\documentclass[aps,prd,floatfix,superscriptaddress,twocolumn,showpacs,amssymb]{revtex4-1}

\usepackage{epsfig}
\usepackage{graphicx}
\usepackage{dcolumn}
\usepackage{bm}

\newcommand{\be}{\begin{equation}}
\newcommand{\ee}{\end{equation}}
\newcommand{\bea}{\begin{eqnarray}}
\newcommand{\eea}{\end{eqnarray}}
\newcommand{\hf} {\frac{1}{2}}
\newcommand{\nn}{\nonumber\\}

\def\eq#1{(\ref{#1})}
\def\la{\langle}
\def\ra{\rangle}

\def\ord#1{{\cal O}\left(#1\right)}

\def\ih{\frac{i}\hbar}
\def\t#1{{\tilde{#1}}}
\def\c#1{{\cal{#1}}}
\def\h#1{{\hat{#1}}}
\def\b#1{{\bar{#1}}} 

\begin{document}
\title{Some consequences of GUP induced ultraviolet wavevector cutoff in one-dimensional
Quantum Mechanics
}

\author{K. Sailer}
\affiliation{Department of Theoretical Physics, University of Debrecen,
P.O. Box 5, H-4010 Debrecen, Hungary}

\author{Z. P\'eli}
\affiliation{Department of Theoretical Physics, University of Debrecen,
P.O. Box 5, H-4010 Debrecen, Hungary}

\author{S. Nagy}
\affiliation{Department of Theoretical Physics, University of Debrecen,
P.O. Box 5, H-4010 Debrecen, Hungary}
\affiliation{MTA-DE Particle Physics Research Group, P.O.Box 51, H-4001 Debrecen, Hungary}

\date{\today}

\begin{abstract}
A projection method is proposed to treat the one-dimensional Schr\"odinger equation
for a single particle when the Generalized Uncertainty
 Principle (GUP) generates an ultraviolet (UV) wavevector cutoff.  The existence of a
unique coordinate representation called the naive one is derived from the one-parameter 
family of discrete coordinate representations.
In this bandlimited Quantum Mechanics  a continuous potential is reconstructed from
discrete sampled  values observed by means of   a particle in maximally localized states.
 It is shown that bandlimitation
  modifies the speed of the center and the spreading time of a Gaussian wavepacket
 moving 
in free space. Indication is found that GUP accompanied by bandlimitation
may cause departures of the low-lying energy levels
of a particle in a box from those in ordinary Quantum Mechanics
much less suppressed than commonly thought when GUP without bandlimitation is in work.

\end{abstract}



\maketitle
\section{Introduction}\label{intro}

There are several theoretical indications  that Quantum Gravity may have consequences in
the behaviour of low-energy quantum systems  \cite{Mead1964,foam,string,Tiwar2008}. The corresponding effective
Quantum Mechanics is believed to be based on the Generalized Uncertainty Principle (GUP) \cite{Magg1994,Kempf199612,Bambi2008,Adler2010,Hossa2010,Noza2011,Hosse2012}, 
various modifications of  Heisenberg's 
Uncertainty Principle (HUP). 
 Recently even
a proposal has been  put forward to probe experimentally   the departure from HUP 
in a direct manner \cite{Piko2011}. Among the various realizations of GUP
 there is a class when 
the deformation of the commutator relation for the operators of the coordinate $\h{x}$ and
 the canonical momentum $\h{p}_x$ depends only on the canonical momentum,
\bea\label{gup}
  \lbrack \h{x}, \h{p}_x\rbrack &= &i\hbar f( \alpha |\h{p}_x|)
\eea
with   the deformation function $f(|u|)$, where $u=\alpha p_x$ and $\alpha =\ord{ \ell_P/\hbar}$ with the 
Planck length $\ell_P\approx 1.616\times 10^{-35}$ m is a small parameter. In the
 present paper we shall restrict ourselves to   deformation functions
for which there exists a  minimal wavelength, i.e. a maximal magnitude $K$ of the wavevector  but the
canonical momentum can take arbitrary large values (as opposed to the cases discussed e.g. in \cite{Ching2012}). Then physical states are restricted to those of finite band width, i.e.
the wavevector  operator $\h{k}_x$ can only take values in the interval 
$\lbrack -K, K\rbrack$.

The mathematical structure of Quantum Mechanics with finite band width is rather 
 delicate \cite{Kempf1993}, even in the one-dimensional case. Namely, while  both operators of the
 wavevector $\h{k}_x$ and the canonical momentum $\h{p}_x$ are self-adjoint, the
 coordinate operator $\h{x}$  cannot be self-adjoint, but only Hermitian symmetric
in order to satisfy the deformed commutator relation \eq{gup}. It has, however, a one-parameter family of self-adjoint 
extensions with eigenvalues determining equidistant grids on the coordinate axis, defining 
a minimal length scale $a$, while the grids belonging to the various extensions
are continuously shifted to each other. This is a sign that positions can only be observed
with a maximal precision $\Delta x_{min}\approx a$, while 
 momenta can be measured with arbitrary accuracy \cite{Kempfma1994,Hinr1995,Kempf1996,Deto2002,Bang2006,Dorsch2011, Noza2011, Mend2011}. 
Although  there exist formal  coordinate
 eigenstates even in that case, but -- as opposed to ordinary Quantum Mechanics, --
 they cannot be approximated now
by a sequence of physical states with uncertainty in position decreasing to zero \cite{Kempfma1994}.
A so-called naive coordinate representation can be built up representing the operator 
algebra of
 $\lbrack\h{x}, \h{k}_x \rbrack = i\hbar$ as $\h{x} \Rightarrow x$ and 
$\h{k}_x\Rightarrow  
-i\partial_x$ on square-integrable functions of the coordinate $x\in \mathbb{R}$, referred to below as 
coordinate wavefunctions. Then the canonical 
momentum
satisfying the relation in Eq. \eq{gup}
can be represented as
 $\h{p}_x \Rightarrow \alpha^{-1} F^{-1}(-i\alpha \hbar \partial_x) $ where 
 $F^{-1}$ stands for the inverse of the function 
$F(u) =\int_0^u \frac{du'}{f(|u'|)}$ and relates the operators of the  wavevector
and the canonical momentum via $\h{k}_x=(\alpha\hbar)^{-1} F(\alpha \h{p}_x)$. 
Let us note that the commutator in the left-hand side of Eq. 
\eq{gup}  is invariant under the reflection $x\to -x,~p_x\to -p_x$ so that
the deformation function should be an even function of $p_x$ which implies that
both functions $F(u)$ and its inverse are even, as well. Furthermore, $f(0)=1$ has to be required in order to recover HUP in the limit $\alpha\to 0$. Here we shall restrict ourselves to
 particular choices of the deformation function which are
monotonically increasing with raising $|u|$ and
for which the limits $F(\pm \infty)=\pm \alpha \hbar K$ remain finite. There are known
 a few explicit cases of such deformation functions, e.g.  $f=1+\alpha^2 p_x^2$ with $F(u)={\rm{~arc~tan~}}u$ \cite{Kempfma1994,Kempf199604}, $f=\exp(\alpha^2 p_x^2)$ with $F(u)=(\sqrt{\pi}/2){\rm{~erf~}}u$ \cite{Dorsch2011}, and   $f=\exp (\alpha |p_x|)$ with $F(u)=(1-e^{-|u|}) {\rm{~sign~}}u$ used after Taylor expansion in \cite{Ali2009,Ali2011}
and
all implying $\hbar K\approx 1/\alpha$ and $a\approx 
\alpha\hbar \approx \ell_P$.

Coordinate space turns out to exhibit features of discreteness \cite{Das2010} and continuity  at the same
 time like information does \cite{Kempf199709,Kempf199806,Kempf199905,Kempf2007,Kempf2010,Kempf201010,Bojo2011}. The main idea is that space can be thought of a differentiable manifold, but the physical degrees of freedom cannot fill it in arbitrarily dense manner.
It has also been conjectured that degrees of freedom corresponding to structures smaller than the resolvable Planck scale turn into internal degrees of freedom \cite{Kempf199706,Kempf199810,Kempf199811,Kempf199907}.
Although the coordinate wavefunctions introduced in the manner described above do not 
have the simple
probabilistic meaning, yet provide  useful tools to characterize the quantum states of the 
particle which can then be analyzed e.g. in terms of maximally localized states \cite{Hinr1995,Kempf1996,Deto2002,Bang2006}. Below we shall discuss the justification
of that naive coordinate representation for one-dimensional bandlimited Quantum Mechanics
in more detail.

 In the present
 paper we shall concentrate on the solution of the Schr\"odinger equation for the coordinate
 wavefunctions in the case when GUP implies finite band width.
It is well-known that GUP directly affects the Hamiltonian through the modification of the 
canonical momentum and that of the kinetic energy operator $ (\h{p}_x^2/2m)-(\hbar^2
 \h{k}_x^2/2m)$ which can be expanded -- when low-energy states are considered -- 
in powers of the small parameter $\alpha$. This pure GUP effect has been treated in
the framework of the  perturbation expansion using the naive coordinate representation and discussed
in detail for various quantum systems (see  Refs.  
\cite{Kempf199604,Ali2011,Nozaz2005,Das2008,Jana2009,Mirza2009,Noza2010,
Ali2010,Pedra2010,Spren2010,Dasma2011,Pedra201110,Pedra201204,Valta2012,
Tau2012} without the quest of completeness), among others for the particle in a
 box \cite{Noza200507,Ali2009,Ali2010,Pedra2010,Pedra2011,Pedra201110,Ali2011}. 
Treatments in the Bargmann-Fock representation  \cite{Kemp1992,Cele1993,Kempf1994},
and  various path-integral formulations \cite{Kempf199602,Kempf199603,Daspra2012} 
 have been worked out.
 Here we shall take  the viewpoint that the bandlimited Quantum Mechanics is an effective
 theory in the framework of which no quantum fluctuations of wavelength smaller than those
 of the order of the minimal length are possible, i.e. the coordinate wavefunctions should not
 contain Fourier components with wavevectors outside of the finite band with
 $k_x\in \lbrack -K, K\rbrack$. In order to built in this restriction into the Schr\"odinger
 equation we propose to use Hamiltonians operating on the subspace of the 
square-integrable coordinate wavefunctions with finite band width.  To
 ensure this we introduce the projector $\h{\Pi}$ onto that subspace and restrict the solutions of the Schr\"odinger equation to the bandlimited subspace $\c{H}$ of the Hilbert space.
We shall also show that the above mentioned naive coordinate representation of the 
bandlimited Hilbert space $\c{H}$ uniquely exists
and solutions of the bandlimited Schr\"odinger equation automatically reflect the symmetry
that the formulations of the theory on any of the equidistant spatial grids of spacing $a$
exhibit the same physical content. Moreover, we shall discuss the reconstruction of a unique continuous bandlimited potential from sampled values taken on such grids by means of maximally localized states. The bandlimitation will be shown to broaden the peaks and smear out the sudden jumps of the microscopic potential over a region of the Planck scale.
Below we shall apply our
 projection method to determine the free motion of a Gaussian wavepacket as well as the energy shifts of the low-lying stationary states of a particle in a box. The stationary problem shall be treated in the framework of first-order perturbation theory.

It is appropriate to make the following remarks: {\it (i)} As to our viewpoint of removing the UV components of the wavefunctions,
it is rather a naive approach to estimate the additional effects due to the existence of the finite UV cutoff. Our viewpoint would be exact when the coordinate space were discrete, but it is not
although the self-adjoint extensions of the coordinate operator have discrete eigenvalues
forming a grid on the coordinate axis. However, there exists a one-parameter family of 
such extensions and that of the corresponding grids which can be transformed into each other by continuous shifts.
 This indicates that quantum fluctuations of wavelength smaller than the minimal length scale
$a$ cannot be probably excluded completely, but rather should have been treated by more sophisticated methods, like e.g. renomalization group methods. 
Fortunately, in the one-dimensional case the solutions of the bandlimited Schr\"odinger equation in the naive coordinate representation used by us reflect inherently the physical 
equivalence  of any of those grids. Therefore,
 our projection technique  can give a reliable order-of-magnitude estimate of the importance of the additional effect  of the UV cutoff as compared to the pure GUP effect. {\it (ii)} As to our choice of the model, the particle in a box, it is rather a toy model. The pure GUP effect on the low-lying stationary states has already been discussed  and
noticed that the model with precisely given box size $L$ is ill-defined in the sense that a change 
of the box size of the order $\ell_P$, i.e. that of the maximal accuracy $\Delta x_{min}$ of the position determination causes an energy shift of the order $\ord{\ell_P/L}$ as compared to the pure GUP effect of the order $\ord{(\ell_P/L)^2}$ \cite{Pedra201112}. In our 
approach the Hamiltonian operates on the subspace of states with finite band width and transforms the 
originally local potential effectively into a nonlocal one. Projecting out the UV components
 of the potential results in a kind of smearing out the edges of the box in regions of the size
 of $\ord{\Delta x_{min}}$. Our purpose is to determine the additional shift of the 
low-lying energy levels caused by the existence of the finite UV cutoff.
We shall see that this turns out to be of the order
$\ord{\ell_P/L}$, being much more significant than the pure GUP effect, so that in some sense the result of \cite{Pedra201112} will be recovered.  Nevertheless, our approach may be a hint that this kind of energy shift might be the true effect, when the physically realistic box
 with smeared out walls is  considered and modeled by performing the projection which
determines the operation of the Hamiltonian on the bandlimited Hilbert space $\c{H}$.

Our  paper is constructed as follows. In Sect.  \ref{proj}  the projection method is introduced
and the integral kernels for the various projected operators determined. 
For one-dimensional bandlimited Quantum Mechanics a justification of the naive coordinate 
representation is given in Sect. \ref{coreps}. 
A method is given in Sect. \ref{smear} which enables one to reconstruct a bandlimited
 continuous potential from sampled values obtained by means of a particle in maximally
 localized state.
The free motion of 
a Gaussian wavepacket is then discussed in Sect. \ref{mofree} in the framework of 
bandlimited Quantum Mechanics.   In Sect. \ref{potprob} the determination of the shifts of the low-lying energy levels of a particle in a potential is formulated in the framework of
the first-order perturbation theory. The problem of a particle in a box is considered in Sect.
\ref{pinb} as the limiting case of a particle in a square-well potential taking the limit of infinite depth. After setting some notations in Subsect. \ref{gener} and recovering the well-known
result for the pure GUP effect in Subsect. \ref{pure}, in Subsect. \ref{fbwshi} the additional energy shifts of the low-lying levels caused by the existence of the finite UV cutoff are shown to be dominant .  Finally,
the results are summarized in Sect. \ref{summ}. 
Several technical details are given in the Appendix. 
App. \ref{exten} reminds the reader on some mathematics relevant for the self-adjoint extension of the Hermitian symmetric coordinate operator.
In App. \ref{unpwf} the wavefunctions
of the unperturbed system, i.e. those for a particle in the square-well potential  in the framework of usual Quantum Mechanics are derived. The operation of the GUP modified kinetic energy operator on exponential functions is determined in App. \ref{kinexp}. The details of the
evaluation of the additional energy shifts of the low-lying energy levels for  a particle in a box
are presented in Appendices \ref{propot} and \ref{tnnev}. In App. \ref{maxlost} maximally
localized states of a particle  are constructed. Finally, in App. \ref{recconp} the bandlimited
potential reconstructed from sampled values of a Dirac-delta like potential is presented.

\section{Projectors onto the subspace of wavefunctions with finite band width}\label{proj}

Let $\psi(x)$, $\t{\psi}(p_x)$ and $\t{\t{\psi}}(k_x)$ be the wavefunctions of the state $|\psi\ra$ in the naive coordinate, canonical momentum  and wavevector representations, respectively. The scalar product of arbitrary states $|\psi\ra$ and $|\phi\ra$ can be written
as
\bea
 \la \phi|\psi\ra&=& \int_{-\infty}^\infty dx \phi^*(x)\psi(x)=
\int_{-K}^K \frac{dk_x}{2\pi} \t{\t{\phi}}^*(k_x)\t{\t{\psi}}(k_x) \nn
&=&
\int_{-\infty}^\infty \frac{dp_x}{2\pi \hbar f(\alpha|p_x|) } \t{\phi}^*(p_x)
\t{\psi}(p_x)
\eea
and is kept invariant under the transformation
\bea
 \psi(x)&=&\int_{-K}^K \frac{dk_x}{2\pi} e^{ik_x x}\t{\t{\psi}}(k_x),\nn
  \t{\t{\psi}}(k_x)& =& \int_{-\infty}^\infty dx e^{-ik_x x}\psi(x) .
\eea
An arbitrary square integrable function $\psi(x)\in L^2(-\infty,\infty)$ contains
ultraviolet (UV) Fourier components with $|k_x|>K$, as well. In order to ensure that the solutions of quantum mechanical eigenvalue equations, as well as that of the Schr\"odinger
 equation belong to the subspace $L^2_K(-\infty, \infty)$ of wavefunctions with finite band width, any operator  $\h{O}$ of an observable should be projected onto that subspace by an appropriate projector $\h{\Pi}$, $\h{O}\Rightarrow \h{\Pi}\h{O}\h{\Pi}$. The projector  should cut off the UV components of any square integrable function $f(x)\in L^2(-\infty, \infty)$, i.e. for its kernel $\Pi(x,y)$ the relation
\bea
 ( \Pi f )(x)&=& \int_{-\infty}^\infty dy \Pi(x,y) f(y) \nn
&=&\int_{K}^K \frac{dk_x}{2\pi}
e^{ik_x x} \t{\t{f}}(k_x)\in L^2_K(-\infty, \infty)
\eea
should hold that implies
\bea
 \Pi(x,y)&=& \int_{-K}^K \frac{dk_x}{2\pi} e^{ik_x(x-y)}=\frac{\sin \lbrack K(x-y)\rbrack}{\pi (x-y)}.
\eea
It is straightforward to show that $\h{\Pi}$ is a projector satisfying $\h{\Pi}^2=\h{\Pi}$.
When $K\to \infty$ the operation of any Hermitian symmetric operator $\h{O}$ on any state $|\psi\ra$ can be represented as
\bea
\int_{-\infty}^\infty dy O(x,y) \psi(y)&=&
\int_{-\infty}^\infty \frac{dk_x}{2\pi} \t{\t{\psi}}(k_x) \t{\t{O}}_{k_x}e^{ik_x x}
\eea
where the kernel $O(x,y)$ and the formal differential operators $O_x$ and $\t{\t{O}}_{k_x}$ are related as
\bea
 O(x,y)&=&O_x\delta(x-y) =\int_{-\infty}^\infty \frac{dk_x}{2\pi} e^{-ik_x y}
\t{\t{O}}_{k_x} e^{ik_xx}.
\eea
Hermitian symmetry implies $O(x,y)=\lbrack O(y,x)\rbrack^*$ and $\t{\t{O}}_{k_x}=
\lbrack \t{\t{O}}_{-k_x}\rbrack^*$. A few examples are summarized in the table:

\small
\begin{center}
\begin{tabular}{|c|c|}
\hline
& \cr
  $O_x$  &   $\t{\t{O}}_{k_x} $ \cr
& \cr
\hline
&\cr 
 $x^n$ $(n\in \mathbb{N})$  &  $\hf \lbrack (-i\partial_{k_x}^\rightarrow)^n+ (i\partial_{k_x}^\leftarrow)^n\rbrack$  \cr
&\cr
 $ (-i\partial_x)^n$ $(n\in \mathbb{N})$ & $k_x^n$  \cr
&\cr
 $|-i\partial_x|=\sqrt{  (i\partial_x)^2 }$ & $|k_x|$ \cr
&\cr
\hline
\end{tabular}
\end{center}
\normalsize
It is straightforward to show that the symmetrized form of $\t{\t{O}}_{k_x}$
for $O_x=x^n$ with partial derivatives $\partial_{k_x}^\rightarrow$ and 
$\partial_{k_x}^\leftarrow$
 acting to the right and left, respectively,  is in agreement 
with Heisenberg's commutation relation $\lbrack \h{x},\h{k}_x\rbrack=i$.

In Quantum Mechanics with finite band width $K$ the kernel $O(x,y)$ should be projected as
\bea
(\Pi O\Pi)(x,y)&=&\int_{-\infty}^\infty dz \int_{-\infty}^\infty du \Pi(x,z)O(z,u)\Pi(u,y) \nn
&=&
\int_{-K}^K\frac{dk_x}{2\pi} e^{-ik_x y}\t{\t{O}}_{k_x} e^{ik_x x}
\eea
while Hermitian symmetry is preserved, $\lbrack (\Pi O\Pi)(y,x)\rbrack^*=(\Pi O\Pi)(x,y)$. Now, one easily finds the kernel of $\h{\Pi}\h{k}_x^n\h{\Pi}$,
\bea
  (\Pi k_x^n\Pi)(x,y)&=&\int_{-K}^K\frac{dk_x}{2\pi} e^{ik_x(x-y)} k_x^n\nn
&=&(-i\partial_x)^n \Pi(x,y),
\eea
and that of $\h{\Pi} \h{x}^n \h{\Pi}$,
\bea
\lefteqn{
  (\Pi x^n\Pi)(x,y)  }\nn
&=&\int_{-K}^K\frac{dk_x}{2\pi}e^{-ik_x y} \hf \lbrack (-i\partial_{k_x}^\rightarrow)^n+ (i\partial_{k_x}^\leftarrow)^n\rbrack e^{ik_x x}\nn
&=&
\hf (x^n+y^n)\Pi(x,y)
\eea
which imply for the kernels of the functions $f(\h{k}_x)$ and $V(\h{x})$ of operators $\h{k}_x$ and $\h{x}$, respectively,  
\bea\label{rules}
  (\Pi f(k_x)\Pi)(x,y) &=& f(-i\partial_x)\Pi(x,y),\nn
(\Pi V(x)\Pi)(x,y)&=&\hf \lbrack V(x)+V(y)\rbrack \Pi(x,y).
\eea

Making use of the projector $\h{\Pi}$ introduced above, the time-dependent and the stationary Schr\"odinger equations can be  written as
\bea\label{balisch}
  i\hbar \partial_t |\psi\ra &=& \h{\Pi}\h{H}\h{\Pi} |\psi\ra,\\
\label{stsch}
\h{\Pi}\h{H}\h{\Pi} |\psi\ra&=&E|\psi\ra,
\eea
respectively, 
in terms of the projected Hamiltonian $\h{\Pi}\h{H}\h{\Pi}$ when Quantum Mechanics with finite band width is considered. The initial condition for the time-dependent Schr\"odinger equation \eq{balisch} should be a physical state, i.e. also bandlimited. The usage of the 
projected Hamiltonian obviously  ensures
that the state will not contain the UV Fourier components. The rules in Eq. \eq{rules} for projecting functions of operators enable one to make both the kinetice energy and the potential energy piece of the projected Hamiltonian explicit in the naive coordinate representation.

\section{Coordinate representations}\label{coreps}

In this section we show that the naive coordinate representation  for  one-dimensional 
bandlimited Quantum Mechanics formulated by means of the projection technique proposed
 in Sect. \ref{proj}
 is equivalent with any of the discrete coordinate representations based on the complete orthonormal sets of eigenvectors of the various self-adjoint extensions of the coordinate operator.
Moreover, the time-dependent  Schr\"odinger equation \eq{balisch} and the stationary one,
Eq. \eq{stsch} in the naive coordinate representation as well as their solutions reflect inherently the
$U(1)$ symmetry which reveals itself in the unique physical content of the formulations
of the  theory in terms of the various discrete coordinate representations. 
Therefore, the so-called naive coordinate representation is correct for one-dimensional bandlimited Quantum Mechanics, although the coordinate wavefunction looses its direct probability meaning as opposed to ordinary Quantum Mechanics.

One might accept the usage of the naive coordinate representation with some reservation 
because the coordinate operator $\h{x}$ is not self-adjoint (see App. \ref{exten}).
 Therefore, no coordinate eigenstates exist in the physical domain and the
 introduction of the
 coordinate representation becomes questionable. There exists, however, a one-parameter family of the
self-adjoint extensions $\h{x}_\theta$  of the coordinate operator, labeled by the
 parameter $\theta\in \lbrack 0, a)$. Any of these extensions for fixed $\theta$ exhibits
 an orthonormal complete set of eigenvectors $|x_n^\theta \ra$ in the bandlimited Hilbert 
space $\c{H}$. In the wavevector representation the corresponding eigenfunctions 
$\la k_x|x_n^\theta\ra=\t{\t{\psi}^\theta}_{\!\!\!\! x_n} (k_x)$ with
 $k_x\in \lbrack -K, K\rbrack$  
are given by Eq. \eq{cooeigf} in App. \ref{exten}, whereas the corresponding discrete 
nondegenerate eigenvalues
$x_n^\theta=an+\theta$ form a  grid with spacing $a$ in the one-dimensional space. 
Therefore, any vector $|\psi\ra\in \c{H}$ can be represented as a linear 
superposition of the base vectors $|x_n^\theta\ra\in \c{H}$ for any given $\theta$, 
$|\psi\ra=\sqrt{a}\sum_{n=-\infty}^\infty \la x_n^\theta|\psi\ra |x_n^\theta\ra$,
 so that  discrete coordinate representations $\c{R}_\theta$ of the normalized vectors
 $|\psi\ra\in \c{H}$
 via the vectors $\{\psi_n^\theta= \la x_n^\theta|\psi\ra\}\in \ell^2$ arise
  (the normalization implies $a\sum_{n=-\infty}^\infty |\psi_n^\theta|^2=1$). 
Then the operators $\h{O}$ over the bandlimited Hilbert space $\c{H}$
should be represented by the countably infinite dimensional matrices
 $O_{nn'}^\theta=\la x_n^\theta| \h{O} |x_{n' }^\theta\ra$. Thus a one-parameter
 family of discrete coordinate
 representations $\c{R}_\theta$ is available. 

Using the wavevector representation, one realizes immediately
that the transformation from a discrete coordinate representation $\c{R}_\theta$ to another  one $\c{R}_{\theta'}$ belongs to the $U(1)$ group because $\t{\t{\psi}^{\theta'}}_{\!\!\!\!\!\!x_n}(k_x)=
e^{ik_x (\theta'-\theta) } \t{\t{\psi}^\theta}_{\!\!\!\!\!x_n}(k_x)$. Such a
 transformation means a shift of the spatial grid from  $\{ x_n^\theta=na+\theta\}$
to $\{x_n^{\theta'}=na+\theta'\}$ on which one describes
one-dimensional bandlimited  Quantum Mechanics. Nevertheless, there should be a 
distinction of  Quantum Mechanics discretized on a grid and the case discussed here. In the
 latter case  space reveals discrete and continuous features at the same time, similarly as
 information does \cite{Kempf199706,Kempf199709,Kempf199806,Kempf199810,Kempf199811,Kempf199905,Kempf199907,Kempf2007,Kempf2010,Kempf201010,Bojo2011}.
Namely, the physical content of any of the discrete coordinate representations 
$\c{R}_\theta$ should be identical, i.e. the $U(1)$ group of the transformations among
 them should be a symmetry. Thus, physics contained in the scalar products 
\bea\label{scpr}
\la \phi|\psi\ra&=&a\sum_{n=-\infty}^\infty \phi_n^{\theta *} \psi_n^\theta
\eea
 of arbitrary vectors $|\psi\ra,~|\phi\ra\in \c{H}$ should be independent of the 
particular choice of the representation $\c{R}_\theta$.
In that respect it is important to underline that the Schr\"odinger equations obtained by using the projection method are given in their abstract forms in Eqs.
\eq{balisch} and \eq{stsch} without referring to any representation, and their solutions $|\psi\ra$ are authomatically in the bandlimited  Hilbert space, $|\psi\ra\in \c{H}$.

Now we show that to each vector  $|\psi\ra\in \c{H}$, i.e. to each vector
$\{ \psi_n^\theta =\la x_n^\theta|\psi\ra\}$ of the representation $\c{R}_\theta$ 
one can associate  a single coordinate wavefunction $\psi(x)$, and the
 latter is independent of the choice of the representation $\c{R}_\theta$ used for its
 construction.
\begin{enumerate}
\item To any position $x\in \mathbb{R}$ belongs exactly a single eigenvalue $x_{n'}^{\theta'}=an'+\theta'=x$ of a particular self-adjoint extension $\h{x}^{\theta'}$ of the coordinate operator (see the discussion at the end of App. \ref{exten}). This enables one to 
construct a wavefunction of continuous variable,
\bea\label{samphil}
  \psi (x) = \la x_{n'}^{\theta'} =x|\psi\ra= \sum_{n=-\infty}^\infty   \la  x_{n'}^{\theta'} =x|x_n^\theta \ra \psi_n^\theta
\eea
in a reliable manner, where $\psi(x_n^\theta)=\psi_n^\theta$ are sampled values of the
coordinate wavefunction on the arbitrarily chosen grid $\{ x_n^\theta\}$. Here figures the
 matrix element $  \la  x_{n'}^{\theta'} =x|x_n^\theta \ra =a\Pi(x,x_n^\theta)$ of the
 unitary transformation from the discrete coordinate representation characterized by 
$\theta$ to the one characterized by $\theta'$, which is directly related to the projector
 $\h{\Pi}$ 
onto the bandlimited Hilbert space. Thus Eq. \eq{samphil} recasted into the form
\bea\label{samphil2}
   \psi (x) =a \Pi(x, x_n)\psi_n^\theta
\eea
can be interpreted as the particular case of the sampling theorem in the bandlimited
Hilbert space  \cite{Kempf199905}, a generalization of Shannon's sampling theorem
\cite{Shan49}.
With a similar logic identifying $x=x_{n'}^{\theta'}$ and $y=x_{n''}^{\theta''}$ as  the $n'$-th and $n''$-th
eigenvalues of the appropriate self-adjoint extensions $\h{x}_{\theta'}$ and $\h{x}_{\theta''}$, respectively,  one is enabled to reexpress any matrix element $(\Pi O\Pi)(x,y)=\la x|
\h{\Pi}\h{O}\h{\Pi}|y\ra$ of an arbitrary operator $\h{O}$ in terms of the matrix $O_{nm}^\theta=\la x_n^\theta|\h{\Pi}\h{O}\h{\Pi}|x_m^\theta\ra=\la x_n^\theta|\h{O}|x_m^\theta\ra $ as 
\bea
   O(x,y)&=& a^2 \sum_{n,m=-\infty}^\infty\Pi(x,x_n^\theta) O_{nm}^\theta \Pi (x_m^\theta, y)
\eea
implying $O(x_n^\theta, x_m^\theta)=O^\theta_{nm}$. Therefore, to any vector of the bandlimited Hilbert space represented by the vector $\{ \psi_n^\theta\}\in \ell^2$ in the discrete coordinate representation $\c{R}_\theta$ there corresponds a continuous coordinate wavefunction $\psi(x)$, and to any operator $\h{\Pi}\h{O}\h{\Pi}$ mapping the bandlimited Hilbert space into itself, i.e. to any matrix of the discrete coordinate representation $\c{R}_\theta$ corresponds the kernel $O(x,y)$ of the continuous coordinate representation.

\item Being aware of the construction of the coordinate wavefunction $\psi(x)=\sum_{n=-\infty}^\infty \la x=x_n^{\theta'}|\h{\openone}|\psi\ra$ via Eq. \eq{samphil}, one can insert any of the decompositions of the unit operator $\h{\openone}=\sum_{n=-\infty}^\infty |x_n^\theta\ra\la x_n^\theta\ra$ over $\c{H}$ associated with any of the representations $\c{R}_\theta$. This means, on the one hand, that the resulting wavefunction $\psi(x)$ associated in the naive coordinate representation to the bandlimited vector $|\psi\ra\in \c{H}$ is unique,
independent of the representation $\c{R}_\theta$ used to its construction. On the other hand it also means that taking sampled values on various spatially shifted grids $\{x_n^\theta\}$ and inserting those into Eq. \eq{samphil2} one obtains finally the same wavefunction of the continuous variable $x$. Similar arguments lead to the unique kernel $O(x,y)$ associated to the operator $\h{O}$ mapping $\c{H}$ into itself.

\end{enumerate}
Thus one can conclude that the solutions of the bandlimited Schr\"odinger equations
for one spatial dimension,
being bandlimited themselves, satisfy the generalized sampling theorem expressed in Eq. 
\eq{samphil2} authomatically and therefore reflect inherently the $U(1)$ symmetry
under the continuous shifts of the spatial grid determined by the discrete eigenvalues of the various self-adjoint extensions of the coordinate operator.

\section{Bandlimited potentials}\label{smear}

Let us here illustrate the manner  one could observe the formally local potential $V(x)$ in the framework of Quantum Mechanics with finite band width. The best one can do to construct
maximally localized states $\varphi_{\b{x}}(x)$ centered at arbitrary positions $\b{x}$
\cite{Hinr1995,Kempf1996,Deto2002,Bang2006}
 and detect the potential exerted on it.  One should however be aware of the fact that  there 
exist only a countable set of physically distiguishable positions $x_n^\theta$, those of the eigenvalues  of an arbitrarily chosen self-adjoint extension $\h{x}^\theta$ of the coordinate operator, which form a grid of spacing $a=\pi/K$. Thus we can sample the potential only at the grid points. Let  $\{|\varphi_n^\theta\ra   \}$ be the sequence of the maximally localized states
centered on the grid points.
Thus, potential values  
\bea\label{smeared}
   \b{V}_n^\theta&=&  \la \varphi_{n}^\theta| \h{\Pi}\h{V}\h{\Pi}|\varphi_{n}^\theta\ra
\eea
can only be observed at a discrete set of points of a grid with spacing $a$.  The maximally 
localized states $|\varphi_{n}^\theta\ra$ belong to the subspace of bandlimited 
wavefunctions, $\h{\Pi} |\varphi_{n}^\theta\ra=|\varphi_{n}^\theta\ra$, so that Eq. 
\eq{smeared} reduces to
\bea\label{smeared2}
  \b{V}_n^\theta &=& \la \varphi_{n}^\theta| \h{V}|\varphi_{n}^\theta\ra.
\eea
According to Shannon's basic sampling theorem on bandlimited
 real functions \cite{Shan49}, a continuous potential $\b{V}(\b{x})$ with finite bandwidth $K$ $(k_x\in \lbrack -K, K\rbrack)$ can perfectly be reconstructed from its values 
$\b{V}_n^\theta=\b{V}(x_n^\theta)$ taken on the set of equidistant  points 
$\{ x_n^\theta\}$ spaced $\frac{2\pi}{2K}=a$ apart:
\bea\label{potrec}
  \b{V}(\b{x}) &=&a \sum_{n=-\infty}^\infty \b{V}_n^\theta \Pi(\b{x}-x_{n}^\theta).
\eea

Now one has to show that the reconstructed potential $\b{V}(\b{x})$ is bandlimited and 
does not depend on the particular choice of the grid, i.e. that of the self-adjoint
extension of the coordinate operator.
One can choose the sampled values $\b{V}_n^\theta$ on the grid 
$\{x_n^\theta=na+\theta\}$
 shifted by any constant $\theta \in \lbrack 0, a)$.
Let $\c{V}(\b{x})$ be the function which takes the values
 $\c{V}(x_n^\theta)=\b{V}_n^\theta$ for any $n\in \mathbb{Z}$ and 
$\theta\in\lbrack 0,a)$. It is generally not bandlimited, implying
\bea
  \c{V}(x)&=& \int_{-\infty}^\infty \frac{dl}{2\pi} e^{ilx} \t{\c{V}}(l).
\eea
Let us now reconstruct a potential $\b{V}(\b{x})$ from the sampled values  $\b{V}_n^\theta$ for given $\theta$ and ask how far the resulted function depends on the particular choice of $\theta$, i.e. that of the particular choice of the self-adjoint extension of the coordinate operator $\h{x}$. According to the reconstruction formula in Eq \eq{potrec}
we get
\bea
  \b{V}(\b{x}) &=& a \sum_{n=-\infty}^\infty \c{V}(x_n^\theta) \Pi(\b{x}-
   x_n^\theta) 
\eea
which implies
\bea
\lefteqn{
  \b{V} (\b{x}+\theta) }\nn
&=& a \sum_{n=-\infty}^\infty \c{V}(na+\theta) 
 \Pi(\b{x}-  na) 
\nn
&=&a \int_{-\infty}^\infty  \frac{dl}{2\pi} \t{\c{V}}(l)\int_{-K}^K \frac{dq}{2\pi}
e^{il\theta+iq\b{x}} \sum_{n=-\infty}^\infty e^{i(l-q)na}.
\eea
Making use of the sum \eq{sumdef}, one obtains
\bea
  \b{V} (\b{x}+\theta) &=&
\int_{-K}^K \frac{dl}{2\pi} \t{\c{V}}(l) e^{il(\b{x}+\theta)} 
=\c{V}_K(\b{x}+\theta)
\eea
i.e. the reconstructed function
$  \b{V} (\b{x}) =\c{V}_K(\b{x})$ which is independent of the choice of the particular self-adjoint representation of $\h{x}$ and is bandlimited, $\c{V}_K(\b{x})=
(\h{\Pi}\c{V})(\b{x})$. One cannot, of course, reconstruct the function $\c{V}(\b{x})$
which contains modes outside of the band $\lbrack -K, K\rbrack$. Below we shall take the 
sampled potential values on the grid $x_n=na$ (for $\theta=0$) and, for the sake of simplicity,  suppress the upper indices $\theta$.

\begin{center}
\begin{figure}[ht]
\epsfig{file=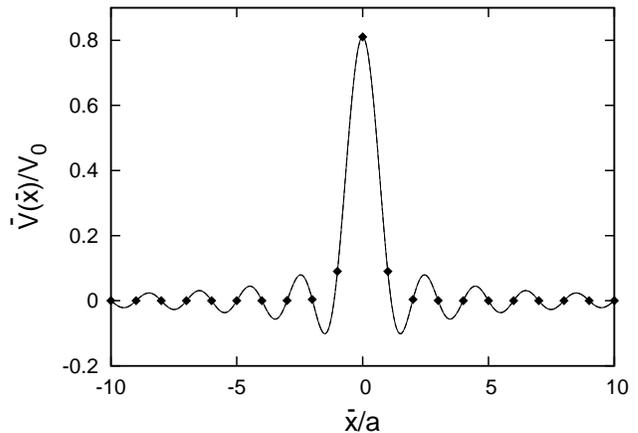,width=6cm,angle=-90}
\caption{\label{fig:delta}
Sampled values (boxes) and the reconstructed bandlimited continuous potential (solid line)
for $V(x)=V_0a\delta(x)$. The numerical result obtained by terminating the sum in Eq. 
\eq{potrec}  at $n=\pm 50$ and the analytic one given by Eq. \eq{didere} cannot be
 distinguished on the figure.
}
\end{figure}
\end{center}

In App. \ref{maxlost} we have determined 
the maximally localized state $\t{\t{\varphi}}_{\b{x}}(k_x)$ in the wavevector representation by making use of 
the method of Detournay, Gabriel, and Spindel \cite{Deto2002}, see Eq. \eq{minposun}.
Rewriting it into the coordinate representation, one finds
\bea\label{maxlocc}
  \varphi_{\b{x}}(x)&=&\int_{-K}^K \frac{dk_x}{2\pi} e^{ik_x x} \t{\t{\varphi}}_{\b{x}}(k_x)\nn
&=&
\sqrt{\frac{a}{2}} \lbrack \Pi(x-\b{x}+\hf a) + \Pi(x-\b{x}-\hf a) \rbrack,
\eea
implying a position inaccuracy of $\la \varphi_{\b{x}}|\h{x}^2-\b{x}^2|\varphi_{\b{x}}\ra=a^2/4$. The wavefunction \eq{maxlocc} is real, has a maximum at
$x=\b{x}$, varies slightly in the small interval $x\in \c{I}_{\b{x}}\equiv\lbrack \b{x}-\hf a, \b{x}+\hf a\rbrack$ centered at the point $\b{x}$
and falls off rapidly in an oscillatory manner outside of the interval $\c{I}_{\b{x}}$. 
 Obviously such a state
enables one to detect a bandlimited potential $\b{V}(\b{x})$ smeared out as compared to $V(x)$.
For example, the Dirac-delta like  potential, $V(x)=V_0 a\delta (x)$ is observed as
a broadened one with finite height $V_0$ and finite width,
\bea\label{didesa}
\b{V}_n&=& V_0 a|\varphi_{x_n}(0)|^2=
  \frac{V_0 a^2}{2} \lbrack \Pi(x_n-\hf a) + \Pi(x_n+\hf a) \rbrack^2.\nn
\eea
The reconstructed bandlimited continuous potential is then given by Eq. \eq{didere2}
in  App. \ref{recconp}  and is shown in Fig. \ref{fig:delta}.
 It is peaked taking values of the order $V_0$ in the interval $\b{x}\in
\lbrack -\hf a, \hf a\rbrack$ and falls off rapidly in an oscillatory manner outside of that interval. The characteristic wavelength of the oscillations is $2a$.

\begin{center}
\begin{figure}[ht]
\epsfig{file=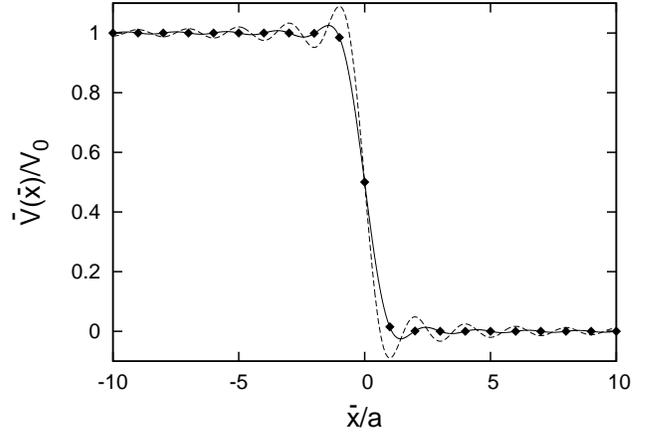,width=6cm,angle=-90}
\caption{\label{fig:theta}
Sampled values (boxes), the reconstructed bandlimited continuous potential (solid line)
and its analytic approximation (dashed line) for the potential step $V(x)=V_0 \Theta(-x)$.
The numerical result (solid line) has been obtained by terminating the sum in Eq. 
\eq{potrec}  at $n=\pm 50$.
}
\end{figure}
\end{center}

Another example is that the finite jump at $x=0$ of the  potential step  $V(x)= V_0\Theta(-x)$ becomes smeared out. The sampled values taken at $x_n=na$ are
\bea\label{samste}
  \b{V}_n&=& V_0\int_{-\infty}^0 dx |\varphi_{na}(x)|^2.
\eea
The values $\b{V}_n$ are monotonically decreasing with increasing $n$.  Moreover, 
the relation $|\varphi_{na}(x)|^2=|\varphi_{0}(x-na)|^2$ holds for the integrand,
implying
\bea
  \b{V}_n&=& V_0\int_{-\infty}^{-na} dx |\varphi_{0}(x)|^2.
\eea
Because $\varphi_0(x)$ is even and normalized to 1, one gets $\b{V}_0=\hf V_0$,
and $\b{V}_{\pm n}= V_0 (\hf \pm r_n)$ with $r_n=\int_0^{na} dx  |\varphi_{0}(x)|^2$ for $n>0$, where trivially $r_n$ increases strictly monotonically from $r_0=0$ to $1$
with $n$ going to infinity so that $\b{V}_n$ takes the value $V_0$ for large negative index $n$ and decreases strictly monotonically through the value $\b{V}_0=\hf V_0$ at $n=0$ to zero when $n$ increases to infinitely large integer values. So long the localized state $\varphi_{na}(x)$ almost entirely extends over a region where the potential is $V_0$ or zero, it detects its original value, but when it `feels' the sudden jump of the potential step  it detects monotonically decreasing values when the increasing sequence $x_n$  runs through
$x=0$. The sampled values already smear out the sudden jump over a region of width $a$.
Reconstruction of the bandlimited potential $\b{V}(\b{x})$ from the sample shall make the fall of the potential around $x=0$ oscillatory.  Instead of evaluating analytically the reconstructed potential itself we shall  illustrate the oscillations introduced by the reconstruction on a more simple example. For that purpose let us choose the sample $\c{V}_n= \c{V}(x_n=an)$ taken of the simple step function $\c{V}(x)=V_0\Theta(-x)$ for which the well-known
integral representation 
\bea
  \c{V}(x)&=& \frac{V_0}{2\pi i}\int_{-\infty}^\infty dl \frac{ e^{-ilx}}{l-i\epsilon}
\eea
holds. The bandlimited potential reconstructed from this sample $\c{V}_n$ is then given as
\bea
  \b{\c{V}}(x) &=& \frac{V_0}{2\pi i}\int_{-K}^K dl \frac{ e^{-ilx}}{l-i\epsilon}.
\eea
Let us evaluate this integral by closing the straightline section $\lbrack -K, K\rbrack$ on the real axis via a half circle $\c{C}_-$ of radius $K$ on the lower half of the complex $l$ plane,
along which one has $l= Ke^{-i\alpha}$ with $\alpha$ running from zero to $\pi$,
\bea
    \b{\c{V}}(x) &=& \frac{V_0}{2\pi i}\int_{\c{C}_-} dl \frac{ e^{-ilx}}{l}
\nn
&=& \frac{V_0}{2\pi } \int_0^\pi d\alpha e^{-iKx\cos \alpha -Kx\sin \alpha}.
\eea
Changing the integration variable from $\alpha$ to $\beta=\alpha-\hf \pi$, one can recast
the integral in the form
\bea
   \b{\c{V}}(x) &=&\frac{V_0}{\pi} \int_0^{\pi/2} d\beta \cos( Kx\sin \beta) e^{-Kx\cos\beta}\nn
&=& V_0\biggl( \hf - \frac{{\rm{~Si~}} (Kx) }{\pi}\biggr).
\label{thetaan}
\eea
The reconstructed potential tends to $V_0$ for $x\to -\infty$, zero for $x\to +\infty$
taking the value $\hf V_0$ at $x=0$ and it falls in an oscillatory manner from the value $V_0$ to zero.  The reconstructed potential and its analytic approximation in Eq. \eq{thetaan} 
are demonstrated in Fig. \ref{fig:theta}.

\section{Motion of a wavepacket in free space}\label{mofree}

The Schr\"odinger equation for the free motion of a particle in one-dimensional space,
say the $x$ axis is 
\bea\label{tdsch}
  i\hbar \partial_t \psi (x,t)&=& \int_{-\infty}^\infty dy (\Pi H^{free}\Pi )(x,y)
\psi(y,t)
\eea
with the Hamiltonian
\bea
  H^{free}(x,y)&=& \frac{1}{2m\alpha^2} \lbrack F^{-1}(-i\alpha \hbar \partial_x)\rbrack^2 \delta(x-y).
\eea
The Hamiltonian $\h{H}^{free}$, the operators of the wavevector $\h{k}_x$ and the canonical momentum $\h{p}_x=F^{-1}(\alpha \hbar \h{k}_x)/\alpha$ are pairwise commuting, and their common set of eigenfunctions are the plane waves $e^{ik_x x}$. The projector
acts on plane waves as
\bea
  \int_{-\infty}^\infty dy \Pi(x,y) e^{ik_xy}& =& \Theta (K-|k_x|)e^{ik_xx}
\eea
so that the subset of plane waves with wavevector $k_x\in \lbrack -K, K\rbrack$, canonical momentum $p_x(k_x)=F^{-1}(\alpha \hbar k_x)/\alpha\in (-\infty, \infty)$ and energy
$E_{k_x}= \lbrack F^{-1}(\alpha \hbar k_x)\rbrack^2/(2m\alpha^2)\in \lbrack 0, \infty)$ eigenvalues, respectively, 
 form 
the set of common eigenfunctions of the operators $\h{\Pi}\h{H}^{free}\h{\Pi}$,
$\h{\Pi}\h{k}_x\h{\Pi}$ and $\h{\Pi}\h{p}_x\h{\Pi}$ in bandlimited Quantum Mechanics. For canonical momenta much less than the Planck momentum, i.e. for $ |p_x|\ll (1/\alpha)$, one can expand the deformation function in powers of the magnitude of $u=\alpha p_x$ as
$f(u)=1+f_1|u|+f_2u^2+\ldots$ which implies the expansions
$ F(u)= u-\hf f_1 u|u|+\frac{1}{3}(f_1^2-f_2) u^3+\ldots$, $F^{-1}(v)=v+\hf f_1 v|v|+\frac{1}{6}(f_1^2+2f_2)v^3+\ldots$ with $v=\alpha \hbar k_x$ and
the well-known modification of the energy eigenvalue due to the GUP
\be
E_{k_x} \approx \frac{\hbar^2 k_x^2}{2m} \biggl( 1
 +f_1\alpha \hbar |k_x| +\frac{1}{12}(7f_1^2+8f_2)\alpha^2 \hbar^2 k_x^2
+\ldots\biggr)
\ee
for $|k_x|\ll K$. As to the plane waves the only effect due to the finite band width is the
lack of plane waves with wavevectors from the UV region.

Now let us seek  the solution of the time-dependent Schr\"odinger equation \eq{tdsch}
in the form of a wavepacket, i.e. that of a superposition from plane waves belonging to the finite band of wavevectors,
\bea
\psi(x,t)&=& \int_{-K}^K \frac{dk_x}{2\pi} a(k_x,t)e^{ik_x x},\\
\label{ateq}
i\hbar \partial_t  a(k_x,t)&=& \frac{\lbrack p_x(k_x)\rbrack^2}{2m} a(k_x,t).
\eea
In order to choose an initial condition $a_0(k_x)=a(k_x,t=0)$ consistent with the generalized uncertainty relation
implied by Eq.  \eq{gup}, one has to be rather cautious. 
In the wavevector representation  the coordinate operator  $\h{x}$ is represented by the
 formal differential operator $i \partial_{ k_x}$
 defined in the dense domain $\c{D}_x\subset L^2[-K,K]$ of
 square integrable wavefunctions $\t{\t{\psi}}(k_x)$, consisting of  absolutely continuous 
functions whose derivatives also belong to $L^2[-K,K]$. The physical domain
$\c{D}_{phys}$ consisting of the wavefunctions satisfying the generalized uncertainty
 relation  should be a subset of the domain $\c{D}_x$, $\c{D}_{phys}\subseteq \c{D}_x$.
We should choose the initial condition from that physical domain, $a_0(k_x)\in \c{D}_{phys}$. For $\h{x}$ being a  symmetric operator,   only two types of boundary
 conditions are allowed \cite{Riesz}. In the set of functions satisfying $\t{\t{\psi}}(K)=C\t{\t{\psi}}(-K)$ with $|C|=1$ the coordinate operator is self-adjoint, but in our case this should be 
excluded because in that case the coordinate operator were diagonal in the coordinate representation and  the GUP in Eq. \eq{gup} could not be satisfied. Thus the wavefunctions
$\t{\t{\psi}}(k_x)\in \c{D}_{phys}$ should satisfy the other type of boundary conditions,
namely Dirichlet's boundary conditions ${\t{\t{\psi}}}(-K)=\t{\t{\psi}}(K)=0$ for which $\h{x}$ is not self-adjoint.
Therefore  we  choose an initial condition $a_0(k_x)$ satisfying the boundary conditions
$a_0(\pm K)=0$, 
\bea\label{inco}
 a_0(k_x)&=& \c{N}
\exp \biggl( - \frac{ \lbrack p_x(k_x)-\b{p}\rbrack^2}{4\sigma_p^2} \biggr),
\eea
corresponding to a Gaussian wavepacket with the mean wavevector   $\b{k}$, the mean canonical momentum $\b{p}=p_x(\b{k})$, the variance $\sigma_p$, and the normalization factor $\c{N}$. Without loss of generality one can assume $\b{k}>0$ (i.e. $\b{p}>0$) and $\sigma_p\ll K\hbar $. The solution of Eq. \eq{ateq} with the initial condition \eq{inco} is
\bea
a(k_x,t)&=&a_0(k_x)\exp \biggl( - \ih \frac{ \lbrack p_x( k_x)\rbrack^2}{2m} t\biggr)
\eea
and the evolution of the wavepacket is given by the time-dependent wavefunction
\be\label{tdwf}
  \psi(x,t)=
\int_{-K}^K \frac{dk_x}{2\pi}a_0(k_x)\exp \biggl( - \ih \frac{ \lbrack p_x( k_x)\rbrack^2}{2m} t+ik_x x\biggr).\nn
\ee

For the sake of simplicity let us assume that the deformation function $f(u)$ is analytic at $u=0$.
For sufficiently sharp distribution, $\sigma_p\ll K\hbar $ one can expand the exponent at the mean $\b{k}$ as
\bea
{\rm{~Exp.~}}&=&-\ih\frac{\b{p}^2}{2m} t+i\b{k}x -A(t)s^2 +iB(x,t)s
\eea
where
\bea
 && A(t)=\frac{\hbar^2 \b{f}^2}{4\sigma_p^2} +\frac{i\beta_2t}{2m\hbar},~~
B(x,t)=x-\frac{\beta_1 t}{2m\hbar},\nn
 &&\beta_1=2\hbar \b{p} \b{f},~~
\beta_2=\hbar^2 \b{f} ( \b{f}+2\alpha^2 \b{p}^2 \b{f}'), 
\eea
with $s=k_x-\b{k}$, $\b{p}=p_x(\b{k})$, $\b{f}=f(\alpha^2 \b{p}^2) >1$,
$\b{f}'=f'(\alpha^2 \b{p}^2) $, $f'(u)=df(u)/du$. The time-dependent wavefunction
\eq{tdwf} can then be rewritten as
\bea
  \psi(x,t)&=&\c{N}
\exp \biggl( -\ih \frac{\b{p}^2}{2m}t +i\b{k}x\biggr) \nn
&&\times \int_{-K+\b{k}}^{K+\b{k}} \frac{ds}{2\pi}
e^{- A(t)s^2+iB(x,t)s}.
\eea
Let us now make use of $\sigma_k\ll K$ and consider the limiting cases (i) $\b{k}=0$
and (ii) $\b{k}=K$. As a good approximation we can replace the definite integrals by 
improper ones  for case (i) $\int_{-K}^K dk_x\Rightarrow \int_{-\infty}^\infty dk_x$
and for case (ii) $\int_{-2K}^0 dk_x \Rightarrow  \int_{-\infty}^0 dk_x$. Because of
the inequality ${\rm{~Re~}}A(t)>0$, the integral remains convergent and one ends up
 with
\bea
  \psi(x,t)&\propto & A^{-\hf}(t) 
\exp \biggl( -\ih \frac{\b{p}^2}{2m}t +i\b{k}x
-\frac{B^2(x,t)}{4A(t)}
\biggr) 
\eea
implying the Gaussian spatial distribution
\bea\label{codis}
|\psi(x,t)|^2 &\propto&
A^{-1}(t) \exp \biggl( -\frac{( x-\b{v}t)^2}{2\sigma_x^2(t)} \biggr),
\eea
centered at the position $\b{x}=\b{v}t$ at time $t$ and moving with the speed $\b{v}=\b{f}\b{p}/m
>\b{p}/m$. Therefore, the center of the wavepacket moves with a larger speed in the
 bandlimited case than in ordinary Quantum Mechanics. The variance of the position distribution is given by
\bea
  \sigma_x^2(t)&=&\frac{|A(t)|^2}{{\rm{~Re~}}A(t)}
= \frac{\hbar^2\b{f}^2}{4\sigma_p^2}\biggl\lbrack 1+ \frac{t^2}{\tau^2}\biggr\rbrack,
\eea
where the characteristic time for the spreading of the wavepacket is given as
\bea\label{spre}
 \tau=\tau_0\frac{\b{f} }{ \b{f}+2\alpha^2 \b{p}^2 \b{f}'  }<\tau_0
\eea
in terms of the characteristic spreading time  $\tau_0=m\hbar/(2 \sigma_p^2)$ in 
ordinary
Quantum Mechanics. We see that the finite band width can cause a much faster spread
of the wavepacket when  its mean wavevector approaches the limiting value $K$.
  
\section{Stationary states of a particle in a potential}\label{potprob}

Let us discuss now the problem of stationary states of a particle in a potential $V(x)$. Even if
 GUP does not imply a finite wavevector cutoff, it results in the modified kinetic energy 
operator $\h{H}^{free}$, introduced in the previous section, so that the Hamiltonian can
 be written as $\h{H}=\h{H}^{free}+\h{V}$. In order to determine the low-energy states
 the non-degenerate stationary perturbation expansion has been widely applied with 
an expansion  in powers of the small parameter $\alpha$ (see e.g.
\cite{Kempf199604,Ali2011,Nozaz2005,Das2008,Jana2009,Mirza2009,Noza2010,
Ali2010,Pedra2010,Spren2010,Dasma2011,Pedra201110,Pedra201204,Valta2012,
Tau2012,Noza200507,Ali2009,Ali2010,Pedra2010,Pedra2011,Pedra201110,Ali2011}). 
When additionally even a finite band width
 is enforced by GUP, the projected Hamiltonian $\h{\Pi}(\h{H}^{free}+\h{V})\h{\Pi}$ 
figures in the stationary Schr\"odinger equation \eq{stsch}.  Now we shall use the 
perturbation expansion for the low-energy states in  another manner, without expanding in 
the small parameter $\alpha$. Namely, we shall simply say that the whole GUP effect, 
including the effect of projection to states with finite band width, is a perturbation and  
account it for  in the first order. Therefore we split the projected Hamiltonian as
\bea
  \h{\Pi}\h{H}\h{\Pi}&=& \h{H}_0 +\h{H'},
\eea
where $\h{H}_0=\h{T}_0+\h{V}$ is the Hamiltonian in ordinary Quantum Mechanics
 with the usual kinetic energy operator
$\h{T}_0=\hbar^2\h{k}_x^2/(2m)$ and
\bea
\h{H'}&=&\h{h}+\h{v}+\h{t}
\eea
 represents the perturbation caused by GUP and the restriction to finite band
 width. The latter consists of the pure GUP effect $ \h{h}=\h{H}-\h{H}_0=\h{H}^{free} - \h{T}_0$ discussed widely in the literature and the pieces $\h{v}=\h{\Pi}\h{V}\h{\Pi}
-\h{V}$ and
$\h{t}=\h{\Pi}\h{H}^{free}\h{\Pi} - \h{T}_0$ responsible for the additional modification
of the potential and the kinetic energy operator due to the restriction of the states to those 
with finite band width. Obviously, the projection alters the local potential and kinetic energy
  into nonlocal quantities.

 Let $\{ \varphi_n\}$ be the complete set of eigenstates of the unperturbed Hamiltonian $\h{H}_0$, 
\bea\label{stschup}
  \h{H}_0 |\varphi_n\ra&=& \epsilon_n |\varphi_n\ra,
\eea
then the energy levels $E_n=\epsilon_n+\Delta \epsilon_n$ of the perturbed system are shifted by
\bea
  \Delta \epsilon_n&=&\la \varphi_n |\h{H}'|\varphi_n\ra
= h_{nn}+v_{nn}+t_{nn},
\eea
where $h_{nn}=\la\varphi_n|\h{h}|\varphi_n\ra$, $v_{nn}=
\la\varphi_n|\h{v}|\varphi_n\ra$, and $t_{nn}=
\la\varphi_n|\h{t}|\varphi_n\ra$ represent the energy
 shifts caused by pure GUP effect and by the projection of the potential and that of the 
kinetic energy, respectively.

\section{Toy model: particle in a box}\label{pinb}

\subsection{Energy shifts of stationary states}\label{gener}
We shall apply the method described in the previous section to a toy model, a particle
bounded in a square-well potential
\bea\label{sqwpot}
  V(x)&=& V_0\lbrack \Theta(-x)+\Theta (x-L)
\eea
of width $L$  and let go finally the depth of the potential well $V_0$ to infinity.
Although a sudden jump of the potential is unrealistic in bandlimited Quantum Mechanics
as emphasized in \cite{Pedra201112}, but in our treatment that problem shall be cured
by projection that makes  potential edges effectively smeared out over a range of the order
 $a$, as argued in Sect.  \ref{smear} previously.

  When GUP effects
 (including finite band width) are neglected,  the solutions $\varphi_n(x)$ 
corresponding to the unperturbed bound states with energy $\epsilon_n<V_0$ in the square well 
potential are given in App. \ref{unpwf}. The expressions for asymptotically large depth $V_0$
of the potential, i.e. for states with $\epsilon_n/V_0\ll1$ are also given. 
 The matrix elements
$h_{nn}$, $v_{nn}$ and $t_{nn}$ contributing additively to the energy shift can be expressed in terms of the various pieces  $\varphi_i(x)$ $(i=I,~II,~III)$ of the wavefunction defined in the intervals $I_i$, respectively (c.f. App \ref{unpwf}). 
 Since the operator $\h{h}$ is local, while the operators
$\h{v}$ and $\h{t}$ are nonlocal due to the projection operator, we can write their matrix 
elements in the form:
\bea
  h_{nn} &=&\sum_{i=I}^{III} h_i,~~h_i=\int_{I_i} dx \varphi_i^*(x)h_x \varphi_i(x),\nn
  v_{nn}&=& \sum_{i,j=I}^{III} v_{i,j},~~v_{i,j}=\int_{I_i} dx\int_{I_j}dy 
 \varphi_i^*(x) v(x,y) \varphi_j (y),\nn
 t_{nn}&=& \sum_{i,j=I}^{III} t_{i,j},~~t_{i,j}=\int_{I_i} dx\int_{I_j}dy 
 \varphi_i^*(x) t(x,y) \varphi_j (y),\nn
\eea
where 
\bea
 h_x&=& \frac{1}{2m}\biggl( \frac{1}{\alpha^2} \lbrack F^{-1}(-i\alpha \hbar \partial_x)\rbrack^2 - (-\hbar^2 \partial_x^2) \biggr),\nn
v(x,y)&=& \hf \lbrack V(x)+V(y)\rbrack \lbrack \Pi(x,y)-\delta(x-y)\rbrack,\nn
t(x,y)&=&\frac{1}{2m\alpha^2} \lbrack F^{-1}(-i\alpha \hbar \partial_x)\rbrack^2 
\lbrack \Pi(x,y)-\delta(x-y)\rbrack \nn
\eea
 are the appropriate kernels. Hermitian symmetry of the
operators $\h{h}$, $\h{v}$, and $\h{t}$, reflection symmetry of the potential to $x=L/2$
and being the operator $\lbrack F^{-1}(\alpha \hbar \h{k}_x)\rbrack^2$ even lead to the symmetry relations
\bea
 && h_{I}=h_{III},\nn
&&v_{j,i}=(v_{i,j})^*,~~v_{III,III}=v_{I,I},~~v_{III,II}=\pm v_{I,II},\nn
&&t_{j,i}=(t_{i,j})^*,~~t_{III,III}=t_{I,I},~~t_{III,II}=\pm t_{I,II}.
\eea 
Here the $\pm$ signs correspond to eigenstates characterized by the wavevectors $k_\pm$.
Furthermore, $v_{II,II}=0$ trivially because $V(x)=0$ for $x\in I_{II}$.

\subsection{Shift due to pure GUP effect}\label{pure}
We call pure GUP effect the energy shift  $h_{nn}$ of stationary states because of the
 modification of the canonical momentum and that of the kinetic energy from $\hbar \h{k}_x$ and $(\hbar \h{k}_x)^2/(2m)$ in ordinary Quantum Mechanics to $\alpha^{-1}
F^{-1}(\alpha \hbar \h{k}_x)$ and $(2m\alpha^2)^{-1} \lbrack F^{-1}(\alpha \hbar \h{k}_x)\rbrack^2$ when GUP is in work.
 Making use of the results of App. \ref{kinexp}  one finds that
the functions $\varphi_i(x)$ $(i=I,~,II,~III)$ are eigenfunctions of the kinetic energy operator. Then one gets
\bea
   h_{II} &=& \frac{1}{2m} \biggl( \frac{1}{\alpha^2} \lbrack F^{-1}(\alpha \hbar k_\pm)
\rbrack^2 -\hbar^2 k_\pm^2 \biggr) \int_{I_{II}}dx |\varphi_{II}(x)|^2,\nn
h_I&=&\frac{1}{2m} \biggl( \frac{1}{\alpha^2} \lbrack F^{-1}( -i\alpha \hbar \kappa)
\rbrack^2+ (\hbar \kappa_\pm)^2 \biggr)\int_{I_I}dx |\varphi_{I}(x)|^2\nn
&=&h_{III}
\eea
and the energy shift due to pure GUP effect
\bea
 h_{nn}&=&
\Biggl\lbrack ( 1+ \frac{\sin k_\pm L}{k_\pm L} ) \biggl( \lbrack F^{-1}(\alpha \hbar k_\pm)
\rbrack^2 - (\alpha \hbar k_\pm)^2 \biggr) \nn
&&
+ \frac{|\rho_\pm|^2}{2\kappa_\pm L} \biggl(\lbrack F^{-1}( -i\alpha \hbar \kappa_\pm)
\rbrack^2+ (\alpha \hbar \kappa_\pm)^2 \biggr)\Biggr\rbrack\nn
&&\times
\Biggl\lbrack  2m\alpha^2 \biggl( 1+ \frac{\sin k_\pm L}{k_\pm L} + \frac{|\rho_\pm|^2}{2\kappa_\pm L} \biggr)\Biggr\rbrack^{-1}.
\eea
Let us consider now the limit $\kappa_\pm \to \infty$, the case of a particle in a box.
In that limit $|\rho_\pm|^2\sim k_\pm^2/\kappa_\pm^2$ and, consequently, one obtains finite energy shift if and only if  the limit $\lim_{\kappa\to \infty} \lbrack F^{-1}( -i\alpha \hbar \kappa)\rbrack^2 \kappa^{-3}=C_\infty$ remains finite. We shall assume that only such deformation functions are physically reasonable for which $C_\infty=0$, which
means that the tails of the wavefunction in the outer regions $I_I$ and $I_{III}$ of the square-well potential give vanishing contributions to the kinetic energy when the depth of the potential becomes infinite. The deformation function $f=1+\alpha^2 p_x^2$ with $F^{-1}(u)=\tan u$ satisfies that
condition because the limit $\lim_{u\to \pm \infty} {\rm{tanh~}}u =\pm 1$ is finite.
Finally, we end up with the pure GUP energy shift $ h_{nn}= R_h \epsilon_n $ with
\bea
  R_h&=&\biggl(\frac{  F^{-1}(\alpha \hbar k_\pm)}{\alpha \hbar k_\pm }\biggr)^2 -1
\eea
for any  deformation function being reasonable in the above discussed sense. For highly
 excited states characterized by wavevectors $k_\pm\approx K$ close to the UV cutoff the
 ratio $R_h$ explodes which signals simply that the perturbation expansion seases to work.
For low-lying states for which our approach is applicable, the expansion in the small parameter $\alpha \hbar k_\pm= n \alpha \hbar\pi/L$ yields
\bea
   R_h &\approx &  \frac{\alpha \hbar \pi}{L}  \biggl\lbrack f_1 n 
+ \frac{1}{12} \biggl( 7f_1^2+8f_2\biggr) \frac{\alpha \hbar \pi}{L} n^2+ \ldots \biggr\rbrack .
\eea
For $h \alpha=\ell_{P}$ and $f=1+\alpha^2p_x^2$ one gets $f_1=0$ and $f_2=1$ and
the ratio
\bea
 R_h&\approx&\frac{2}{3}  \biggl(\frac{\ell_P}{2L}\biggr)^2 n^2+\ord{ (n\ell_P/L)^2}
\eea
raising quadratically with increasing $n$ and being independent of the mass of the particle
 in the box. Thus we recovered the result obtained in Refs. \cite{Noza2006} and 
\cite{Pedra2011} (given after Eq. (14) for $j=1$).   An order-of-magnitude estimate gives
$(\ell_P/L)^2 \approx 10^{-40},~10^{-50}$, and $10^{-58}$ for boxes of the size
 $L=10^{-15}$ m (size of a nucleon),  $10^{-10}$ m (size of a H-atom), and $10^{-6}$ m (the wavelength of infrared radiation), respectively. So even for the first few thousands
of energy levels the pure GUP correction remains a tiny correction.

\subsection{Shift due to finite band width}\label{fbwshi}
Finite band width, i.e. the existence of the  finite UV wavevector cutoff $K$ results in the absence of  quantum fluctuations with UV wavevectors $|k_x|>K$, that is expressed in our approach by the projection of the operators of potential and kinetic energies. The energy shift
$v_{nn}$ of the $n$-th energy level caused by the replacement of the potential by its projected counterpart can be expressed in terms of the independent integrals $v_{I,I}$, $v_{I,II}$ and $v_{I,III}$ when the symmetry relations discussed above are accounted for. Here $v_{I,I}$ and $v_{I,III}$ are real because $\varphi_I(x)$ and $\varphi_{III}(x)$ are real functions.
As discussed in App. \ref{propot} the leading order contribution comes from the integral
$v_{I,II}$ in the limit of infinite potential depth, while the other independent integrals $v_{i,j}$ are suppressed like powers of $1/\kappa_{\pm}$ as compared to it. One finds (c.f. Eq.
\eq{vnn}) that $v_{nn}$ vanishes for the energy levels $n$ even and for the energy levels
$n$ odd it is given as (c.f. Eqs. \eq{v12}, \eq{ic+}, and \eq{vnn})
\bea
 v_{nn}&\approx &\frac{\hbar^2k}{2mL}\biggl(\frac{2k}{K\pi }\lbrack 4\sin^2 (\nu\pi) +(n\pi)^2 \cos(2\nu \pi)\nn
&&+\ord{n^4}\rbrack
+\ord{ (k/K)^2}\biggr).
\eea
The vanishing of $v_{nn}$ for even $n$ is a consequence of the particular form of the wavefunction in a square-well potential, namely the alternating sign of the tail of the wavefunction in the region $III$ with the alteration of even and odd $n$ values in the numeration of the 
stationary states with increasing energy.
The ratio of the energy shift $v_{nn}$ for odd $n$ to the unperturbed energy $\epsilon_n$ of the stationary state $n$,
\bea
 R_v &=&\frac{v_{nn}}{\epsilon_n} \approx \frac{2}{KL\pi}\lbrack 4\sin^2 (\nu\pi)
+(n\pi)^2\cos(2\nu \pi)\nn
&&+\ord{n^4}\rbrack+\ord{ (\ell_P/L)^2}
\eea
turns out to take values of the order $(\ell_P/L)$.

Thus,  the potential energy shift due to the UV cutoff seems to be many orders of magnitude
larger -- at least for the lowest energy levels -- as compared to the energy shift caused by
 pure GUP, because it holds $R_h/R_v\approx \ord{\ell_P/L}$.  It is remarkable that $R_v$ oscillates strongly  with the
 fine-tuning of the length of the box $L$ confining the particle.  The variation of $\nu$ in the interval $\lbrack 0,1\rbrack$ corresponds to the tiny 
change of the box size $L$ in a range of $\ord{\ell_P}=\ord{a}$, the size of the grid
 constant, as well as that of the maximal accuracy $\Delta x_{\min}$ of position
 determination. 
 Therefore, an averaging over $\nu$
might be more reliable when one wants to incorporate the indefiniteness of the size of the box, a direct consequence of the impossibility to determine positions more precisely than the
distance $\Delta x_{min}$. This yields then
\bea\label{avrv}
\int_0^1 d\nu R_v&\approx & \frac{4}{KL\pi}\lbrack 1+\ord{n^4}\rbrack+\ord{(KL)^{-2}}.
\eea
 An order-of-magnitude estimate gives
$(KL)^{-1}\sim (\ell_P/L) \approx 10^{-20},~10^{-25}$, and $10^{-29}$ for boxes of the size
 $L=10^{-15}$ m (size of a nucleon),  $10^{-10}$ m (size of a H-atom), and $10^{-6}$ m (the wavelength of infrared radiation), respectively. These are still small effects but $20$ to $30$ orders of magnitude larger than the energy shift due to pure GUP without wavevector cutoff.

Another contribution $t_{nn}$ arises due to finite band width, which represent the difference of the expectation values of the projected and unprojected kinetic energy operators. According to the symmetry relations, the only independent integrals contributing to the kinetic energy shift $t_{nn}$ are $t_{I,I},~t_{II,I},~t_{III,I},~t_{II,II}$ as given in Eqs. \eq{tpro} in
App. \ref{tnnev}. Among them $t_{II,II}$ is the only one surviving the limit $\kappa\to \infty$, that of taking the square-well potential with   infinite depth. According to Eqs. \eq{t22a} and the estimate in Eq. \eq{cikapp} one obtains
\bea\label{t22b}
 t_{II,II}&\approx & R_t (1+R_h)\epsilon_n \approx R_t \epsilon_n
\eea
with  
\bea
  R_t (\nu)&\approx & \frac{t_{II,II} }{\epsilon_n} \approx \c{I}_K(L/2)-1.
\eea
Here $R_t(\nu)$ is the ratio defined as  the additional shift of the expectation value of the kinetic energy due to the finite band width divided by the unperturbed energy of the stationary state $n$. While the additional energy shift $v_{nn}$ appeared to be vanishing for states with even $n$, the shift $t_{nn}$ is nonvanishing for all $n$.
As seen in Eq. \eq{cikapp} this ratio oscillates with the fine tuning of the length $L$ of the potential box again. Referring to the impossibility of determining positions and distances more precisely than $\Delta x_{min}$ in bandlimited Quantum Mechanics, one can perform averaging over $\nu\in \lbrack 0,1\rbrack$, similarly to the case for the shift of the expectation value of the potential operator due to the finite band width.
The averaging over the interval $\nu\in \lbrack 0,1\rbrack$ results in
\bea
\int_0^1 d\nu R_t(\nu) &\approx & \frac{4}{KL\pi} +\ord{(KL)^{-2}},
\eea
being the same as the relative contribution of the shift of the potential energy in Eq. \eq{avrv}.

\section{Summary}\label{summ}
The free motion of a wavepacket and the energy levels of a particle in a box have been
discussed in one-dimensional Quantum Mechanics when the existence of a finite band
width, i.e. that of an UV wavevector cutoff $K$ is present as the consequence of the GUP.
The latter is implemented by generalizing  Heisenberg's commutation relations for 
quantization with  the particular
choice of the deformation function $f(|u|)$ $(u=\alpha p_x)$  occuring in the commutator relation for the coordinate
$\h{x}$ and the canonical momentum $\h{p}_x$. Deformation functions $f(|u|)$ being strictly monotonically increasing with $|u|$, for which $F(u) =\int_0^u \frac{du'}{f(|u'|)}$
remains finite in the limits $u\to \infty$, provide such UV cutoff, $F(\pm \infty)=\pm \alpha \hbar K$.  We took the point of view that the 
existence of the UV wavenumber cutoff, corresponding to infinite canonical momentum
eigenvalue, excludes the UV components of the wavefunction. In order to enforce this in the 
naive coordinate representation, the Hamiltonian and all operators of observables should be
sandwiched by a projector restricting wavefunctions to the subspace of bandlimited
wavefunctions. Such a projector $\h{\Pi}$ has been constructed and the rules for the operators acting on the band-limited subspace have been established. It has also been shown
that the proposed projection method justifies the usage of the naive coordinate representation through a generalization of Shannon's basic sampling theorem taken from information theory
to one-dimensional bandlimited Quantum Mechanics. It has been discussed the relation of 
the naive coordinate representation using coordinate wavefunctions of continuous variable 
to discrete coordinate representations based on the self-adjoint extensions of the coordinate
 operator. 

A method is proposed to observe potential values exerted on a particle by means of 
preparing it in a state of maximal localization. Although any self-adjoint
 extension of the coordinate operator enables one to take such sampled values at the discrete
points of equidistant spacing $a=\pi/K$, the reconstruction of a bandlimited continuous 
potential is possible according to Shannon's basic sampling theorem on bandlimited real functions. Applying that reconstruction procedure the broadening  the Dirac-delta like potential
and the smearing out the potential step over a region of the order of the spacing $a$ have been shown, both accompanied with oscillations of wavelength of ca. $2a$, as well.

It has been shown that the free motion of the wavepacket is modified as the consequence 
of the finite band width. The center of a Gaussian 
wavepacket with mean canonical momentum $\b{p}$ moves with a speed $\b{V}$ larger than 
$\b{v}=\b{p}/m$ (c.f. Eq. \eq{codis}) and the spreading time $\tau$ of the wavepacket gets smaller than the corresponding characteristic time $\tau_0$ in usual Quantum Mechanics
(c.f.  Eq. \eq{spre}).  The ratios $\b{V}/\b{v}$ and $\tau/\tau_0$ are strictly monotonically increasing and decreasing functions of the mean momentum $\b{p}$, respectively.

The shifts of the low-lying energy levels of a particle in a box have been determined 
considering the effect of GUP and the additional effect caused by the UV cutoff in first-order
perturbation theory. For the pure GUP effect the well-known result has been recovered
being of the order $(\ell_P/L)^2$ for the box size $L$ (in terms of the Planck length $\ell_P$) and the deformation function $f=1+\alpha p_x^2$. 
The additional effect caused by the UV cutoff, i.e. by the projection to the subspace of states with finite band width, has occurred as the shift of the expectation values of both the potential and  kinetic energy operators and turned out to be of the order $\ell_P/L$. This result indicates that the effect of GUP on low-energy Quantum Mechanics may be much more significant
 indirectly, through  the existence of the finite UV cutoff than directly by providing small correction terms to the Hamiltonian. It is also remarkable that the additional effect caused by
the UV cutoff appeared to have an oscillatory dependence on the variation of the box size $L$ in a range of the minimal accuracy of position determination. We have suggested
 that any observation of the box size should average over such a range principally, therefore
 the true correction should be averaged over that range, as well.

%

\section*{Acknowledgements}
The work is supported by the TAMOP-4.2.2/B-10/1-2010-0024 project.
The project is co-financed by the European Union and the European Social
Fund.

\appendix

\section{Self-adjoint extensions of the coordinate operator}\label{exten}

Here we give a short summary of some mathematical results \cite{Riesz} relevant for the
 self-adjoint extension of the coordinate operator (c.f. also \cite{Kempf1993,Kempf199709,Kempf199806,Kempf199905,Kempf2007,Kempf2010,Kempf201010,Bojo2011}).
In the wavevector representation the states are represented by the wavevector 
wavefunctions
$\t{\t{\psi}}(k_x)$, the coordinate operator by the formal differential operator 
$i\partial_{ k_x}$ (it is called formal because the definition of an operator should include
 the boundary conditions set on the functions on which it operates, as well).
The coordinate operator $\h{x}$ is defined in the dense domain 
$\c{D}(\h{x})\subset L^2\lbrack -K,K\rbrack$ of those
 square integrable functions, which are  absolutely continuous (implying infinite 
differentiability) inside the interval $\lbrack -K, K\rbrack$, and whose derivatives also
 belong to $L^2[-K,K]$. The physical domain
$\c{D}_{phys}$ consisting of the wavefunctions satisfying the generalized uncertainty
 relation implied by GUP in Eq. \eq{gup}
 should be a subset of the domain $\c{D}(\h{x})$, $\c{D}_{phys}\subseteq \c{D}(\h{x})$.
In order to have real expectation values, the coordinate operator 
$\h{x}$ should be  symmetric. Therefore,  only two types of boundary conditions are
allowed to be set at $k_x=\pm K$ for the wavefunctions $\t{\t{\psi}}(k_x)$.
 It is well-known  that the operator $i\partial_{k_x}$ is self-adjoint for the
 boundary conditions $\t{\t{\psi}}(K)=C\t{\t{\psi}}(-K)$ with $|C|=1$, i.e.
 $C=e^{i\alpha}$ with $\alpha\in \lbrack 0, 2\pi )$, and it is not
self-adjoint although Hermitian symmetric operating on functions satisfying Dirichlet's boundary conditions 
${\t{\t{\psi}}}(-K)=\t{\t{\psi}}(K)=0$.
If $\h{x}$ were self-adjoint in the physical domain $\c{D}_{phys}$,
its eigenstates  $e^{-ik_x x}$ would belong to that domain. This  cannot, however, be the
 case because the formal eigenfunctions of the operator $i\partial_{k_x}$ 
cannot satisfy the boundary condition $e^{-iKx} =e^{i\alpha+iKx}$ for arbitrary eigenvalue $x$. One can also argue, that would $\h{x}$ be self-adjoint in
the physical domain, the coordinate eigenstates with zero position uncertainty were physical states and that would contradict GUP implying a wavevector cutoff, i.e. a nonvanishing
minimal position uncertainty.  Therefore, it remains the only possibility of $\h{x}$ being symmetric on functions with
Dirichlet's boundary condition. 

In this case the adjoint operator $\h{x}^\dagger $ represented formally also by $i\partial_{k_x}$ has the domain 
 $\c{D}(\h{x}^\dagger)$ consisting of all differentiable functions of $L^2[-K,K]$
whose derivatives also belong to $L^2[-K,K]$ and which are not restricted by any kind of boundary conditions. Thus clearly one has $\c{D}(\h{x})\subset \c{D}(\h{x}^\dagger)$.
Symmetric operators can be characterized by their deficiency indices $\nu_\pm$, the
 dimensions of the nullspaces of the operators $\h{x}^\dagger -(\pm i)\xi$, respectively, 
where $\xi\in\mathbb{R}$. The solutions of the equations
\bea
0
&=&  (i\partial_{k_x} \mp i \xi) \t{\t{\varphi}}_{\pm} ,~~\xi\in \mathbb{R}
\eea
are the square-integrable functions $\t{\t{\varphi}}_{\pm}(x) = e^{\pm \xi k_x}$,
which span  one-dimensional null-spaces, so that 
the deficiency indices of $\h{x}$ are equal, $\nu_+=\nu_-\equiv\nu=1$.
 Then there exist 
self-adjoint extensions $\h{x}_e=\h{x}_e^\dagger$ of $\h{x}$. According to
 the general theory, these can be constructed by  means of the boundary conditions 
prescribed for the functions of the domain $\c{D}(\h{x}_e)$. In the case with 
$\nu=1$ a single boundary condition is needed to specify
the domains $\c{D}(\h{x}_e)$
of the self-adjoint extension $\h{x}_e$ and that can be done in terms of a single function
 $f_1(k_x)\in \c{D} (\h{x}^\dagger)$ being linearly independent relative to 
$\c{D}(\h{x})$ (because $\h{x}$ is a closed operator):
\bea
  \c{D}(\h{x}_e)&=& \{ f |f\in\c{D}(\h{x}^\dagger), \lbrack f(k_x)f_1^*(k_x)
\rbrack^K_{-K}=0,\nn
&&
 \lbrack f_1(k_x)f_1^*(k_x)\rbrack^K_{-K}=0,  f_1 \in \c{D}(\h{x}^\dagger)  \}
\eea
with $\lbrack f g^*\rbrack^b_a=f(b)g^*(b)-f(a)g^*(a)$. The function $f_1$ linearly
 independent relative to $\c{D}(\h{x})$ can be constructed
as the linear combination $f_1(k_x)= \beta h_1(k_x)+\gamma h_{-1}(k_x)$
$(\beta, \gamma\in \mathbb{C})$
of a number of $2\nu=2$ functions, $h_s(k_x)$ $(s=1, -1)$ which are linearly 
independent 
relative to $\c{D}(\h{x})$, i.e. for which  the relations
\bea
  \lbrack h_r h_s\rbrack^K_{-K}&=& r(2K)^2\delta_{rs},~~r,s=1,-1 
\eea
hold. These  are the functions $h_1=K+k_x$ and $h_2=K-k_x$.
Then one finds $f_1(k_x)=\beta (K+k_x)+\gamma (K-k_x)$ and
\bea
 0&=& \lbrack f_1f_1\rbrack^K_{-K}= \beta\beta^* h_1^2(K) - \gamma\gamma^* h_{-1}^2(-K)\nn
&=&
   (2K)^2(\beta\beta^*-\gamma\gamma^*)
\eea
implying
\bea
&& \beta/\gamma=\frac{1}{(\beta/\gamma)^*},~~\Rightarrow~~ |\beta/\gamma|^2=1,\nn
&&\beta/\gamma=e^{i\alpha+2\pi n i},~~\alpha\in[0,2\pi),~n\in \mathbb{Z}
\eea
and
\bea
 &&\lbrack ff_1\rbrack^K_{-K}= f(K)f_1^*(K)-f(-K)f_1^*(-K)=0,
\eea
restricting the domain of the self-adjoint extension to the functions with boundary conditions
\bea\label{thebc}
&& f(K)=f(-K) \frac{f_1^*(-K)}{f_1^*(K)}= f(-K)e^{-i\alpha},
\eea
given through a particular choice of the parameter $\alpha\in \lbrack 0,2\pi)$ (when $f(\pm K)\not=0$) and to those
with  Dirichlet's boundary conditions.
This means, that the various self-adjoint extensions $\h{x}_e$ of the operator $\h{x}$ are
obtained  when the domain of the formal differential operator $i\partial_{k_x}$ is
defined as 
 \bea
  \c{D}(\h{x}_e)&= &\c{D}(\h{x}) \cup \{ f |
 f\in\c{D}(\h{x}^\dagger),~~ f(K)= f(-K)e^{-i\alpha},\nn
&& ~~~~~~~~~~~~\alpha\in[0,2\pi) \}.
\eea
Thus the various self-adjoint extensions are identical on the domain $\c{D}(\h{x})$ and
can be parametrized by the real number $\alpha
\in[0,2\pi)$ so that we can write for them $\h{x}_e=\h{x}_\alpha$.
The eigenfunctions $e^{-ik_x x}\in \c{D}(\h{x}_\alpha)$ (but $\not \in \c{D}(\h{x})$)
of the particular self-adjoint extension $\h{x}_\alpha$ are
those satisfying the boundary conditions in Eq. \eq{thebc}
 \bea
    e^{-iK x}=e^{iKx-i\alpha}
\eea
belonging to the eigenvalues $x=x_n^\theta=na+\theta$ with $a=\pi/K$ and $\theta=\alpha/2K\in \lbrack 0, a)$. These eigenvalues determine a grid of equidistant points 
on the coordinate axis with spacing $a$. We can change the notation of the particular self-adjoint extensions from $\h{x}_\alpha$ to $\h{x}_\theta$.  Since the eigenfunctions
\bea\label{cooeigf}
\t{\t{\psi}^\theta}_{\!\!\!\!\!x_n}(k_x)&=&\sqrt{a} e^{ik_x x_n^\theta},~~~~
k_x\in \lbrack -K, K\rbrack
\eea
 of any
particular self-adjoint extension $\h{x}_\theta$ form a complete orthonormal set,
satisfying the orthonormality conditions (the upper index $\theta$ of the eigenvalues are suppressed) 
\bea\label{xeigort}
 \delta_{nn'}&=& \int_{-K}^K\frac{dk_x}{2\pi}\t{\t{\psi}^\theta}^*_{\!\!\!\!\!x_n}(k_x)
\t{\t{\psi}^\theta}_{\!\!\!\!\!x_n'}(k_x)\nn
&=&
\int_{-\infty}^\infty \frac{dp_x}{2\pi f(\alpha|p_x|)} \t{\psi}^{\theta*}_{x_n}(p_x)
\t{\psi}^\theta_{x_n'}(p_x),
\eea
There exists a one-parameter family of such orthonormal bases in the Hilbert space $\c{H}$ of bandlimited wavefunctions. Moreover, the one-parameter family of all eigenvalues, i.e. the
union $\cup_{\theta\in\lbrack 0,a)} \{ x_n^\theta \}$
of all sets of eigenvalues of the various self-adjoint extensions can be mapped trivially in a 
one-to-one way on the real line $\mathbb{R}$. 
 Namely, each real number $x\in \mathbb{R}$ occurs  as  a single eigenvalue
 of a single self-adjoint extension $\h{x}_\theta$.

\section{Bound states in a square-well potential}\label{unpwf}

The solution $\varphi_n(x)$ with energy $\epsilon_n<V_0$ of the stationary Schr\"odinger equation 
\eq{stschup} for the square-well potential in Eq. \eq{sqwpot} should be sewed from the functions $\varphi_i(x)$ $(i=I,~II,~III)$
defined in the intervals $I_I: x\in(-\infty, 0\rbrack$, $I_{II}: x\in \lbrack 0,L\rbrack$, and
$I_{III}: x\in\lbrack L, \infty)$, respectively,
satisfying the equations
\bea\label{fieqs}
-\frac{\hbar^2}{2m}\varphi_{II}''(x) &=& \epsilon \varphi_{II}(x),\nn
-\frac{\hbar^2}{2m}\varphi_{i}''(x) &=& -(V_0-\epsilon)\varphi_i(x),~~i=I,~III
\eea
(We shall suppress the index $n$ numerating the energy levels.)
The boundary conditions ensure continuous differentiability of the solution at the boundaries $x=0$ and $x=L$ of the various intervals, as well as exponential fall-off at infinities $|x|\to  \infty$ for
square integrability. Looking for the solutions of Eqs. \eq{fieqs} in the form
\bea
&& \varphi_{II}(x)=Be^{ikx}+ Ce^{-ikx},\nn
 && \varphi_I(x) =Ae^{\kappa x},~~\varphi_{III}=De^{-\kappa x},
\eea
with the real parameters $k=\sqrt{2m\epsilon}/\hbar$ and $\kappa=\sqrt{2m(V_0-\epsilon)}\hbar$,
the boundary conditions at $|x|\to \infty$ are automatically satisfied. The boundary conditions at $x=0$ and $x=L$ result in the set of homogenoeus linear equations, 
\bea
\label{aeq}
  A&=&B+C,\\
\label{akaeq}
A\kappa&=&ik(B-C),\\
\label{deq}
Be^{ikL}+Ce^{-ikL}&=&De^{-\kappa L},\\
\label{dkaeq}
ik(Be^{ikL}-Ce^{ikL})&=&-\kappa De^{-\kappa L}.
\eea
There exists a nontrivial solution for the coefficients $A$, $B$, $C$, and $D$ if and only if the determinant of the set of linear equations vanishes, yielding the implicit equation 
\bea
  e^{2ikL}&=& \biggl( \frac{\kappa-ik}{\kappa+ik}\biggr)^2
\eea
for the energy eigenvalues $\epsilon_n=\hbar^2 k_n^2/(2m)<V_0$ of bound states with discrete values  $k_n$ and $\kappa_n=\sqrt{2m(V_0-\epsilon_n)}\hbar$. The numeration of the states by the
integer $n=1,2,\ldots$ can be established in the limit $V_0\to \infty$ when $\kappa\to \infty$ and
$e^{2ikL}\to 1$ which yields the wavevectors $k_n=n\pi/L$ with $n\in \mathbb{N}$. It is straightforward to realize that the wavevectors $k_\pm$ satisfying
\bea
  e^{ik_\pm L}&=&\mp \frac{\kappa_\pm -ik_\pm}{\kappa_\pm+ik_\pm}
\eea
with $\kappa_\pm=\sqrt{2mV_0-k_\pm^2\hbar^2}/\hbar$ in the limit $V_0\to \infty$ behave
as $k_-=2n'\pi/L$ and $k_+=(2n'-1)\pi/L$ for $n'=1,2,\ldots$.

Eqs. \eq{aeq} and \eq{deq} can be used to express $A$ and $D$ via $C$ and $D$, 
whereas taking the ratio of the appropriate sides of Eqs. \eq{aeq} and \eq{akaeq} one finds
after trivial manipulations
\bea
 C&=& -\frac{\kappa_\pm -ik_\pm}{\kappa_\pm+ik_\pm} B=\pm e^{ik_\pm L} B
\eea
implying $A=B\rho_{\pm}$ and $D= \pm Be^{\kappa_\pm L}\rho_\pm$ with $\rho_{\pm}=1\pm e^{ik_\pm L}$. 
The normalization condition $\int_{-\infty}^\infty dx |\varphi_n(x)|^2=\sum_{i=I}^{III}
\int_{I_i} dx |\varphi_i(x)|^2=1$ with
\bea
\int_{I_{II}}dx |\varphi_{II}(x)|^2 &=&  |B|^2 2L \biggl(1+\frac{\sin k_\pm L}{k_\pm L}\biggr),\nn
\int_{I_i}dx |\varphi_{i}(x)|^2 &=& \frac{|\pm B\rho_\pm|^2}{\kappa_\pm},~~i=I,~III
\eea
yields
\bea
  |B|^{-2}&=& 2L \biggl(1+\frac{\sin k_\pm L}{k_\pm L}+\frac{|\rho_\pm|^2}{2\kappa_\pm L} \biggr)\nn
& \approx& 2L\biggl( 1 + \frac{2}{\kappa_\pm L} + \ord{(k/\kappa)^3} \biggr).
\eea

For later use we shall need the coefficients for asymptotically large values of $V_0$, i.e.
those of $\kappa \gg k$ which means $\kappa\gg 1/L$ for sufficiently low lying states. Keeping the
 terms up to the order $\ord{\kappa^{-2}}$ one obtains
\bea\label{asy1}
  \rho_\pm &\approx & \frac{2ik_\pm}{\kappa_\pm}+ \frac{2\kappa_\pm^2}{k_\pm^2}+
\ord{(k/\kappa)^3},\nn
|\rho_\pm|^2&\approx & \frac{4\kappa_\pm^2}{k_\pm^2}+
\ord{(k/\kappa)^3},\nn
\sin k_\pm L&\approx & \pm\frac{2k_\pm}{\kappa_\pm} +
\ord{(k/\kappa)^3},\nn
\eea
and
\bea\label{asy2}
B&\approx & \frac{1}{\sqrt{2L}}\biggl( 1\mp \frac{1}{\kappa_\pm L} +\frac{3}{2\kappa_\pm^2 L^2} + \ord{(k/\kappa)^3} \biggr),\nn
B\rho_\pm &\approx &  \frac{1}{\sqrt{2L}}\biggl( \frac{2ik_\pm}{\kappa_\pm}+\frac{2k_\pm^2}{\kappa^2_\pm}\mp \frac{2ik_\pm}{\kappa_\pm^2L} + + \ord{(k/\kappa)^3} \biggr),\nn
|B\rho_\pm|^2&\approx &\frac{1}{2L}\biggl(\frac{4k_\pm^2}{\kappa_\pm^2} 
+\ord{(k/\kappa)^3} \biggr).
\eea

\section{Kinetic energy operator on exponential functions}\label{kinexp}

In general an arbitrary operator function $g(\h{k}_x)$ operates on a function $f(x)$ as 
\bea
  g(\h{k}_x)f(x)&=& \int_{-\infty}^\infty dy \int_{-\infty}^\infty \frac{dk_x}{2\pi}
   e^{-ik_x y}g(k_x)e^{ik_x x}f(y) \nn
&=&\int_{-\infty}^\infty \frac{dk_x}{2\pi}g(k_x) 
   \t{\t{f}}(k_x)e^{ik_x x}
\eea
with the Fourier transform of the function $f(x)$,
\bea
  \t{\t{f}}(k_x) &=& \int_{-\infty}^\infty dx e^{-ik_x x} f(x)
\eea

In case of exponential functions, however, the Fourier transform is not a well-behaved 
function, rather a distribution. Therefore, one has to be careful when using the integral 
representation of various operators on exponential functions. In order to be more definite,
one can consider the Fourier transforms of exponential functions as limits of Gaussian 
integrals. For  $f(x)=e^{isx}$ $(s\in \mathbb{R})$, one can write
\bea
 \t{\t{f}}(k_x)&=& 
\lim_{\sigma\to 0^+} \int_{-\infty}^\infty dx e^{i(s-k_x) x- \hf \sigma^2 x^2}\nn
&=& \lim_{\sigma\to 0^+} \sqrt{\frac{2\pi}{\sigma^2} } e^{ - \frac{(s-k_x)^2}{2\sigma^2} }
\eea
and similarly for $f(x)= e^{s x}$, $(s\in \mathbb{R})$
\bea
 \t{f}(k_x)&=&  
\lim_{\sigma\to 0^+} \int_{-\infty}^\infty dx e^{-ik_x x+ sx- \hf \sigma^2 x^2}\nn
&=& \lim_{\sigma\to 0^+} \sqrt{\frac{2\pi}{\sigma^2} } e^{ \frac{ (s-ik_x)^2}{2\sigma^2} },
\eea
and the trivial operation of the powers of $\h{k}_x$  on exponential functions,
$ (-i\partial_x)^n e^{isx} =s^n e^{isx}$ and $(-i\partial_x)^n e^{sx}= (-is)^n e^{sx}$, can also be recovered  by saddle point integration:
\bea
  (-i\partial_x)^n e^{isx} &=& \lim_{\sigma\to 0^+} \sqrt{\frac{2\pi}{\sigma^2} } 
\int_{-\infty}^\infty \frac{dk_x}{2\pi} e^{ - \frac{(s-k_x)^2}{2\sigma^2} } k_x^n
e^{ik_x x} \nn
&= &\lim_{\sigma\to 0^+} \frac{1}{\sqrt{2\pi} \sigma } (s+i\sigma^2 x)^n e^{isx-\hf \sigma^2 x^2}\nn
&&\times
\int_{-\infty}^\infty d\eta e^{ -\frac{\eta^2}{2\sigma^2} }
=   s^n e^{isx}
\eea
and
\bea
  (-i\partial_x)^n e^{sx} &=& \lim_{\sigma\to 0^+} \sqrt{\frac{2\pi}{\sigma^2} } 
\int_{-\infty}^\infty \frac{dk_x}{2\pi} e^{ \frac{ (s -ik_x)^2}{2\sigma^2} 
  } k_x^n e^{ik_x x} \nn
&=&
\lim_{\sigma\to 0^+} \frac{1}{\sqrt{2\pi}\sigma } (-is+i\sigma^2 x)^n
e^{-\hf \sigma^2 x^2+ sx}\nn
&&\times
\int_{-\infty}^\infty d\eta e^{-\frac{\eta^2}{2\sigma^2} }
= (-is)^n e^{sx}.
\eea
Then the less trivial action of powers of the operator $|\h{k}_x|$ can be obtained in a similar manner:
\bea
  |-i\partial_x |^n e^{isx}&=& 
 \int_{-\infty}^\infty \frac{dk_x}{2\pi}|k_x|^n 
   e^{ik_x x}2\pi \delta (s-k_x)= |s|^n e^{isx}\nn
\eea
or otherwise
\bea
 |-i\partial_x |^n e^{isx}&=& 
 \lim_{\sigma\to 0^+} \sqrt{\frac{2\pi}{\sigma^2} } \int_{-\infty}^\infty \frac{dk_x}{2\pi}|k_x|^n e^{ik_x x} e^{ - \frac{(s-k_x)^2}{2\sigma^2} }\nn
& = &
 \lim_{\sigma\to 0^+} \frac{1}{\sqrt{2\pi} \sigma} | s+i\sigma^2 x |^n 
e^{isx-\hf \sigma^2 x^2}\nn
&&\times\int_{-\infty}^\infty d\eta e^{ - \frac{\eta^2}{2\sigma^2}}
= |s|^ne^{isx}
\eea
and 
\bea
 |-i\partial_x |^n e^{sx} &=&\lim_{\sigma\to 0^+}\sqrt{\frac{2\pi}{\sigma^2} } 
\int_{-\infty}^\infty \frac{dk_x}{2\pi}|k_x|^n 
   e^{ik_x x+ \frac{(s-ik_x)^2}{2\sigma^2} )}\nn
&=&
 \lim_{\sigma\to 0^+}\frac{1}{\sqrt{2\pi}\sigma } 
|-is+i\sigma^2 x|^n e^{sx-\hf \sigma^2 x^2} \nn
&&\times\int_{-\infty}^\infty d\eta e^{-\frac{\eta^2}{2\sigma^2} }
= |s|^n e^{sx}.
\eea 
The lesson we have learned is the following. Let $ \c{T}(v^2, |w|)$ be
 given as a double  Taylor expansion in powers of $v^2$ and $|w|$, then
the operator obtained by inserting $v=w=-i\hbar \alpha \partial_x$ acts on
exponential functions as
\bea
 &&  \c{T}(-\hbar^2 \alpha^2\partial_x^2, |-i\hbar \alpha\partial_x|) e^{isx}
=  \c{T}( \hbar^2 \alpha^2 s^2, \hbar \alpha |s|)e^{isx},\nn
&&
\c{T}(-\hbar^2 \alpha^2\partial_x^2, |-i\hbar \alpha\partial_x|) e^{sx} =
\c{T}(-\hbar^2 \alpha^2 s^2, \hbar \alpha |s| ) e^{sx}\nn
\eea
for $s\in \mathbb{R}$, or otherwise exponential functions are eigenfunctions of the
operator $\c{T}( -\hbar^2 \alpha^2 \partial_x^2, |-i\hbar \alpha \partial_x|)
= [F^{-1}(\hbar \alpha \h{k}_x)]^2$, i.e.
that of the
 kinetic energy operator. Moreover, the function $\lbrack F^{-1}(u)\rbrack^{2}$
is even, so that the eigenvalues of the kinetic energy operator when acting on exponential functions is independent of the sign of $s$. 

\section{Evaluation of $v_{nn}$}\label{propot}
The independent integrals $v_{i,j}$ contributing to the matrix element $v_{nn}$ are
\bea
\lefteqn{v_{I,I} = |B\rho_\pm|^2 V_0 }\nn
&&\times \biggl\lbrack\int_{-\infty}^0 dx e^{\kappa_\pm x}
\int_{-\infty}^0 dy e^{\kappa_\pm y} \Pi(x-y)
- \frac{1}{2\kappa_\pm}\biggr\rbrack,\nn
\lefteqn{v_{I,III} 
= \pm |B\rho_\pm|^2 V_0 }\nn
&&\times\int_{-\infty}^0 dx e^{\kappa_\pm x}
\int_L^\infty dy e^{\kappa_\pm (L-y)} \Pi(x-y),\nn
\lefteqn{v_{I,II}   = \hf |B|^2 \rho_\pm^* V_0  }\nn
&&\times  \int_{-\infty}^0 dx e^{\kappa_\pm x}
\int_0^L dy (e^{ik_\pm y}\pm e^{ik_\pm (L-y)} )\Pi(x-y).\nn
\eea
We shall determine these integrals in the asymptotic limit $V_0\to \infty$ implying $\kappa \to \infty$. (Let us
 suppress the lower index $\pm$ still it is not disturbing.) Then
the main contribution to the $x$ integral comes in each cases from $x=0$ due to the 
extremely rapidly falling off factors $e^{\kappa x}$ in the integrands. Therefore, we
can expand the slowly variing $x$-dependent factor $\Pi(x-y)$ of the integrand at $x=0$
and recast the integral over $x$ like
\bea
\lefteqn{
  \int_{-\infty}^0 dx e^{\kappa x}\Pi(x-y) }\nn
&\approx &
   \int_{-\infty}^0 dx e^{\kappa x}\lbrack \Pi(-y) + x\Pi'(-y) +\ord{x^2} \rbrack
\nn
&\approx &
   \lbrack \Pi(-y) +\Pi'(-y) \partial_{\kappa} +\ord{\partial_\kappa^2} \rbrack
 \int_{-\infty}^0 dx e^{\kappa x}\nn
&\approx &
    \lbrack \Pi(-y) +\Pi'(-y) \partial_{\kappa} +\ord{\partial_\kappa^2} \rbrack \kappa_\pm^{-1} \nn
&\approx &
 \kappa^{-1}\Pi(-y) +\kappa^{-2} \Pi'(-y) + \ord{\kappa^{-3}}
\eea
with the notations $\Pi'(u)=d\Pi(u)/du$, $\Pi''(u)=d^2\Pi(u)/du^2$.
Thus we get an expansion of the integral in powers of $1/\kappa$. In the limit $V_0\to \infty$, i.e. $k/\kappa\to 0$ one is only interested in the leading order terms of $v_{i,j}$'s.

With similar logic, one can expand the integrals over $y$ in the expressions of
$v_{I,I}$ and $v_{I,III}$ also in powers of $1/\kappa$. Since the even and odd derivatives of $\Pi(u)$ at $u=0$ are finite and zero, respectively, (e.g. $\Pi(0)=K/\pi$, $\Pi'(0)=0$, $\Pi''(0)=-K^3/\pi$) the following well-defined expansion
occurs in the expression of $v_{I,I}$,
\bea
\lefteqn{
  \int_{-\infty}^0 dy e^{\kappa y} \biggl(  \kappa^{-1}\Pi(-y) +\kappa^{-2} \Pi'(-y) + \ord{\kappa^{-3}} \biggr)  }\nn
&\approx &
 \biggl(  \kappa^{-1}\Pi(0) +\kappa^{-2} \Pi'(0) + \ord{\kappa^{-3}} \biggr) \kappa^{-1}\nn
&\approx &
\kappa^{-2} \Pi(0)+\ord{\kappa^{-3}}
\eea
which implies
\bea
  v_{I,I}&\approx & |B\rho|^2 V_0 \biggl(\frac{\Pi(0)}{\kappa^2}-\frac{1}{2\kappa}+\ord{\kappa^{-3}} \biggr)\nn
&\approx & |B\rho|^2 V_0\biggl(-\frac{1}{2\kappa}
 +\ord{\kappa^{-2}}\biggr).
\eea
Similarly, expanding the $y$-dependent factor of the integrand of $v_{I,III}$ at $y=L$,
the leading order term  provides
\bea
 v_{I,III}& =& \pm |B\rho|^2 V_0\biggl(\frac{\Pi(-L)}{\kappa^2}
+\ord{\kappa^{-3}} \biggr)
\eea
in the limit $V_0\to \infty$ (with the sign $\pm$ when the lower index $\pm$ is everywhere restored). Furthermore, the integral $v_{I,II}$ can be recasted as
\bea
\lefteqn{
 v_{I,II}\approx  \frac{1}{2\kappa} |B|^2 V_0 \rho^* }\nn
&&\int_0^L dy (e^{ik y}
\pm e^{ik(L-y)} ) \lbrack  \Pi(-y)+\ord{\kappa^{-1}}
\rbrack \nn
\lefteqn{
\approx 
\frac{1}{2\kappa i} |B|^2 V_0 \rho^*  }\nn
&&\int_{-K}^K\frac{dq}{2\pi}
\biggl( \frac{ e^{i(k-q)L}-1}{k-q} \mp e^{ikL} \frac{e^{-i(k+q)L}-1}{k+q} \biggr)\nn
&\approx &
\frac{1}{2\kappa i 2\pi} |B|^2 V_0 \rho^* 
\biggl( \int_{-KL-kL}^{KL-kl} du \frac{1-e^{-iu}}{u} \nn
&&
\pm e^{ikL}\int_{-KL+kL}^{KL+kL} du \frac{1-e^{-iu}}{u} \biggr)
\eea
for asymptotically large values of $\kappa$. Here the expression in the bracket can be recasted into the sum
\bea
\biggl(\ldots \biggr)&=&
 I_{c-}+iI_{s-}\pm e^{ikL}( I_{c+}+iI_{s+})\nn
&=&
\lbrack \pm e^{ikL}-1\rbrack I_{c+}+i \lbrack 1\pm  e^{ikL}\rbrack I_{s+}
\eea
with 
\bea
  I_{c\pm}&=&\int_{-KL\pm kL}^{KL\pm kL} du\frac{1-\cos u}{u},\nn
  I_{s\pm}&=&\int_{-KL\pm kL}^{KL\pm kL} du\frac{\sin u}{u}.
\eea
and $I_{c-}=-I_{c+}$, $I_{s-}=I_{s+}$.
For the low-lying excited states we can Taylor-expand these integrals 
in the small parameter $k/K\ll 1$,
\bea
    I_{c+}&\approx &  \frac{k}{K} \biggl(4\sin^2\frac{KL}{2} +(kL)^2\cos(KL)
\nn
&&+\ord{(kL)^4}\biggr)+ \ord{(k/K)^2},\nn
I_{s+} &\approx & \int_{-KL}^{KL} du \frac{\sin u}{u}+ \ord{(k/K)^2}.
\eea

Making use of the asymptotic relations $\kappa_\pm\approx \sqrt{2mV_0}/\hbar$, $\pm e^{ik_\pm L}\to -1$ and those in Eqs. \eq{asy1} and \eq{asy2}, one easily establishes the order-of-magnitude relations
$v_{I,I}\sim \ord{ \kappa^{-1}}$, $v_{I,III}\sim \ord{\kappa^{-2}}$, and the leading order contribution comes from 
\bea\label{v12}
  v_{I,II}&\approx&  \frac{\hbar^2 k}{4mL\pi} I_{c+}  ,
\eea
One can write $KL=2\pi(N+\nu)$ where $N$ and $0\le\nu <1$ stand for
the integer  and  fractional parts of the ratio $KL/(2\pi)$, respectively.
Consequently, one finds $\sin^2(KL/2)=\sin^2(\nu \pi)$ and 
\bea\label{ic+}
I_{c+}&=& \frac{k}{K}\lbrack 4\sin^2(\nu \pi)+ (n\pi)^2\cos (2\nu \pi)
+\ord{n^4}\rbrack \nn
&&+ \ord{(k/K)^2}
\eea 
which has the order of magnitude $k/K\sim n(\ell_P/L)$.

Summing the  contributions of leading order in $1/\kappa$, and making use of $v_{III,II}=\pm v_{I,II}$, finally one obtains for the matrix element $v_{nn}$, 
\bea\label{vnn}
  v_{nn}&\approx & v_{I,II}+v_{II,I}+v_{III,II}+v_{II,III}\nn
&\approx & 2(v_{I,II}+v_{III,II})\nn
&\approx &
\Biggl\{ \begin{array}{llll} 4v_{I,II} & {\mbox{for}} & n&   {\mbox{odd}}\cr
0 & {\mbox{for}} & n&   {\mbox{even}} \end{array}
\eea
The vanishing of the potential energy shift for $n$ even is the consequence of the sign difference
of the wavefunction  in the outer regions $I_I$ and $I_{III}$, i.e. that of the coefficients $A$ and $D$ implying $v_{III,II}=-v_{I,II}$ in that case.

\section{Evaluation of $t_{nn}$ }\label{tnnev}
The independent integrals contributing to the kinetic energy shift $t_{nn}$ arising due to
the finite band width are as follows:
\bea\label{tpro}
  t_{I,I}&=& \frac{|B\rho|^2}{2m\alpha^2}\int_{-\infty}^0 dx e^{\kappa x}
  \lbrack F^{-1}(-i\alpha\hbar\partial_x) \rbrack^2\nn
&&\times\biggl(
\int_{-\infty}^0 dy \Pi(x-y)e^{\kappa y}- e^{\kappa x}\biggr),\nn
 t_{II,I}&=&\frac{ |B|^2\rho}{2m\alpha^2} \int_0^L dx (e^{-ikx}\pm e^{-ik(L-x)})
 \lbrack F^{-1}(-i\alpha\hbar\partial_x) \rbrack^2 \nn
&&\times \int_{-\infty}^0 dy \Pi(x-y) e^{\kappa y}
,\nn
t_{III,I}&=& \pm \frac{|B\rho|^2}{2m\alpha^2} \int_{L}^\infty dxe^{\kappa(L-x)}
 \lbrack F^{-1}(-i\alpha\hbar\partial_x) \rbrack^2 \nn
&&\times\int_{-\infty}^0  dy \Pi(x-y)e^{\kappa y}
,\nn
t_{II,II}&=& \frac{|B|^2}{2m\alpha^2} \int_0^L dx (e^{-ikx}\pm e^{-ik(L-x)})
\lbrack F^{-1}(-i\alpha\hbar\partial_x) \rbrack^2 \nn
&&\times\biggl(\int_0^L  dy \Pi(x-y)
(e^{iky}\pm e^{ik(L-y)}) \nn
&&- (e^{ikx}\pm e^{ik(L-x)})\biggr)
.
\eea
Let us expand the integrals over $y$ in $t_{I,I}$, $t_{II,I}$ and $t_{III,I}$ in powers of $\kappa^{-1}$
similarly to what we did in App.  \ref{propot} for the $x$-integrals,
\bea
\lefteqn{
 \int_{-\infty}^0dy \Pi(x-y) e^{\kappa y}         }\nn
&\approx&
   \int_{-\infty}^0dy \lbrack \Pi(x)- y\Pi'(x) +\ord{y^2}\rbrack e^{\kappa y}
\nn
&\approx &
 \lbrack \Pi(x)- \Pi'(x)\partial_\kappa +\ord{\partial_\kappa^2}\rbrack \kappa^{-1}\nn
&\approx &
 \kappa^{-1}\Pi(x) +\ord{\kappa^{-2}} .
\eea
Then a similar $(1/\kappa)$-expansion of the  integrals over $x$ occurring
 in $t_{I,I}$ and $t_{III,I}$, respectively, is possible:
\bea
\kappa^{-1}\int_{-\infty}^0 dx e^{\kappa x} G_\Pi(x) &\approx &
\kappa^{-2}G_\Pi(0) +\ord{\kappa^{-3}},
\nn
\kappa^{-1}\int_{L}^\infty dx e^{\kappa(L-x) }  G_\Pi(x) &\approx &
  \kappa^{-2}G_\Pi(L)+\ord{\kappa^{-3}}
\eea
with $G_\Pi (x)=\lbrack F^{-1}(-i\alpha \hbar \partial_x) \rbrack^2 \Pi(x)$.
Then one finds in the leading order of $1/\kappa$,
\bea
t_{I,I}&\approx &   \frac{|B\rho|^2}{2m\alpha^2}\biggl(
   \kappa^{-2}G_\Pi(0)
-\frac{   \lbrack F^{-1}(-i\alpha \hbar \kappa)\rbrack^2}{2\kappa}
+\ord{\kappa^{-3}} \biggr)\nn
&\approx &\ord{ \kappa^{-3}},\nn
t_{III,I}&\approx &\pm  \frac{|B\rho|^2}{2m\alpha^2}  \lbrack\kappa^{-2}G_\Pi(L)
+ \ord{ \kappa^{-3}} \rbrack\nn
&\approx & \ord{ \kappa^{-4}},
\eea
where we made use of our particular choice of the deformation function $f(u)$.
The leading order terms of $t_{II,I}$ are given as
\bea
 t_{II,I}&\approx & \frac{ |B|^2\rho}{2m\alpha^2} \kappa^{-1}\int_0^L dx (e^{-ikx}\pm e^{-ik(L-x)})G_\Pi(x)\nn
&\approx & \ord{\kappa^{-2}}.
\eea
Therefore the only independent integral contributing to $t_{nn}$ in the limit $\kappa\to \infty$ is $t_{II,II}$.

As to the next we try to estimate the integral $t_{II,II}$ in the limit $\kappa \to \infty$. 
Let $\chi_{\lbrack 0,L\rbrack}(x)$ be the characteristic function of the interval $x\in\lbrack 0,L\rbrack$. 
In order to perform the integral over $y$, let us first rewrite the trivial integral 
$\int_0^L dy
\delta(x-y) e^{iky}=e^{ikx} \chi_{\lbrack 0,L\rbrack}(x)$ as a limit,
\bea
\lefteqn{
   \int_0^L dy\delta(x-y)e^{iky}    }\nn
&=&
\lim_{\Lambda\to \infty}
\int_{-\Lambda}^\Lambda \frac{dq}{2\pi}e^{iqx}  \int_0^L dy e^{-i(q-k)y}\nn
&=&
e^{ikx}\lim_{\Lambda\to \infty}
\int_{-\Lambda}^\Lambda \frac{dq}{2\pi}e^{i(q-k)x}  i\frac{e^{-i(q-k)L}-1}{q-k}\nn
&=& e^{ikx}\lim_{\Lambda\to \infty}
\int_{-\Lambda}^\Lambda \frac{dp}{2\pi}e^{ipx} i \frac{ e^{-ipL}-1}{p}\nn
&=&
 e^{ikx}\lim_{\Lambda\to \infty}
\int_{-\Lambda}^\Lambda \frac{dp}{2\pi}\biggl\lbrack\frac{\sin pL}{p} \cos px
 - \lbrack\cos (pL)-1\rbrack\frac{\sin px}{p} \biggr\rbrack\nn
&=&
e^{ikx}\lim_{\Lambda \to \infty} \hf \int_{-1}^1 ds 
\biggl\lbrack\frac{\sin \lbrack s\Lambda(L-x)\rbrack }{s\pi} + \frac{\sin (s \Lambda x)}{s\pi} \biggr\rbrack\nn
&=&
e^{ikx} \hf \int_{-1}^1 ds \biggl\lbrack \chi_{\lbrack0,L\rbrack} (x) 
\biggl( \delta(s) +\delta (s)\biggr) \nn
&&+ \Theta (L-x) \biggl( -\delta (s) +\delta(s)\biggr)\nn
&&
+ \Theta (-x) \biggl( \delta(x)-\delta(x)\biggr) \biggr\rbrack\nn
&=&
e^{ikx} \chi_{\lbrack0,L\rbrack} (x)  \hf  \int_{-1}^1 ds \biggl(\delta(s) 
+ \delta(s) \biggr)\nn
&=& e^{ikx}\chi_{\lbrack0,L\rbrack} (x).
\eea
Now let us evaluate $\int_0^L dy \Pi(x-y)e^{iky}$ in a similar manner, where the limit
 $\Lambda \to \infty$ is removed and $\Lambda$ replaced by the finite cutoff $K$,
\bea
\lefteqn{
\int_0^L dy \Pi(x-y)e^{iky}               }\nn
&=&
\int_{-K}^K \frac{dq}{2\pi}e^{iqx}  \int_0^L dy e^{-i(q-k)y}\nn
&=&
e^{ikx} \hf \int_{-1}^1 ds 
\biggl\lbrack\frac{\sin \lbrack s K(L-x)\rbrack }{s\pi} 
+ \frac{\sin (s K x)}{s\pi} \biggr\rbrack\nn
&=& e^{ikx} \c{I}_K(x).
\eea
Making use of this and the limit $\pm e^{ik_\pm L}\to -1$ for $\kappa\to \infty$,
one can recast the integral $t_{II,II}$ in the form
\bea
  t_{II,II}&=& -\frac{|B|^2}{2m\alpha^2}\int_0^L dx  (e^{ikx}-e^{-ikx} )\lbrack \c{I}_K(x)-1\rbrack\nn
&&\times
  \lbrack F^{-1}(-i\alpha \hbar \partial_x)\rbrack^2 (e^{ikx}-e^{-ikx} ).
\eea
In order to obtain an order-of-magnitude estimate for $t_{II,II}$,
let us note that the integral $\c{I}_K(x)$ for sufficiently large cutoff $K$  should be a rather
 smooth function of $x$ because it is
independent of $x$ for $K\to \infty$, $\c{I}_{K\to \infty}(x)\to 1$.  Then the following 
approximations seem to be justified:
(i) the replacement of the kinetic energy operator by its eigenvalue when acting on the
 functions $e^{\pm ikx}$ (c.f. App. \ref{kinexp}),
(ii) replacement of $\c{I}_K(x)$ by $\c{I}_K(L/2)$.
The integral over $x$ reduces then to
\bea
  \int_0^L dx\sin^2( kx)& =&
\hf L
\eea
 for $kL=n\pi$ with any $n=1,2,\ldots$ and one finds
\bea\label{t22a}
 t_{II,II}&\approx &
\frac{  \lbrack F^{-1}(\alpha \hbar k)\rbrack^2 }{2m\alpha^2}\lbrack
\c{I}_K(L/2)-1\rbrack
\eea
with
\bea\label{cik}
\c{I}_K(L/2)-1&=&
\frac{2}{\pi} \int_0^{KL/2} du \frac{\sin u}{u} -1,
\eea
where the new integration variable $u=sKL/2$ has been introduced.

Let $N'$ and $\nu'\in \lbrack 0,1)$ the integer and fractional parts of $KL/(4\pi)$, respectively. The integral \eq{cik} can be rewritten as
\bea\label{cikapp}
  \c{I}_K(L/2)-1&=&-\frac{2}{\pi} \int_{KL/2}^\infty du \frac{\sin u}{u}\nn
&\approx &  -\frac{2}{\pi} \frac{2}{KL} \int_{\nu 2\pi}^{2\pi} du \sin u \nn
&\approx &  \frac{4}{KL\pi} \lbrack 1-\cos (2\nu \pi)\rbrack .
\eea
Here one has split the interval $u\in \lbrack KL/2,\infty)$ into 
subintervals $\lbrack (N'+\nu')2\pi, (N'+1)2\pi\rbrack,~\lbrack 2(N'+j)\pi, 
2(N'+j+1)\pi\rbrack$ with $j=1,2,\ldots$ and replaced the factor $1/u$ in the integrand by $\lbrack(N'+\nu')2\pi \rbrack^{-1},~\lbrack (N'+j)2\pi\rbrack^{-1}$ with $j=1,2,\ldots$, respectively in the subsequent intervals.

\section{Maximally localized states} \label{maxlost}
Following the method of Detournay, Gabriel, and Spindel \cite{Deto2002} we construct
the maximally localized state centered at position $\b{x}$ for arbitrary deformation function.
One should look for the state $|\phi \rangle$ with a given undetermined position uncertainty $\mu$, i.e. the state satisfying 
\bea
  \mu^2&=& \la  \phi| \h{x}^2-\b{x}^2 |\phi\ra
\eea
and the subsidiary conditions
\bea\label{xev}
  \la  \phi| \h{x}|\phi\ra=\b{x}
\eea
and $ \la  \phi|\phi\ra=1$. Then one selects out the state  $|\varphi_{\b{x}}\ra$ with the
 minimal value of $\mu$. In the
wavevector representation, the
variational problem is equivalent with the solution of the differential equation
\bea\label{dgseq}
 0&=& \lbrack (i\partial_{k_x})^2 -\b{x}^2 -\mu^2 -2\lambda  (i\partial_{k_x}-\b{x})\rbrack
  \t{\t{\phi}}(k_x),
\eea
where the wavefunction  $ \t{\t{\phi}}(k_x)$ should satisfy Dirichlet's
boundary conditions $\t{\t{\phi}}(\pm K)=0$ and be normalized, $\int_{-K}^K\frac{dk_x}{2\pi} |  \t{\t{\phi}}(k_x)|^2=1$; $\lambda\in \mathbb{R}$
 is a Lagrange-multiplier that should be determined from the subsidiary condition \eq{xev}.
Looking for the solutions in exponential form, one finds two independent solutions, $\t{\t{\phi}}_\pm =e^{ik_x \sigma_\pm}$ with $\sigma_\pm =-\lambda\pm \sqrt{ (\lambda-\b{x})^2+\mu^2}$. Then the general solution of Eq. \eq{dgseq} is given as 
$\t{\t{\phi}}(k_x)=A e^{ik_x\sigma_+}+ Be^{ik_x\sigma_-}$ and one gets from Dirichlet's boundary conditions, $\t{\t{\phi}}(\pm K)=0$, that
\bea
  K(\sigma_+-\sigma_-)=2\sqrt{ (\lambda-\b{x})^2+\mu^2} &=& N\pi ,~~N\in \mathbb{N} \nn
B=-Ae^{iN\pi}.
\eea
Thus one finds  the sets of solutions:
\bea
\t{\t{\phi}}_{N=2n}(k_x) &=& 2iA e^{-i k_x\lambda} \sin \biggl(\frac{n\pi}{K}k_x\biggr),
\nn
\t{\t{\phi}}_{N=2n-1}(k_x) &=&2A e^{-i k_x\lambda} \cos \biggl(\frac{(2n-1)\pi}{2K}k_x\biggr)
\eea
with $n\in\mathbb{N}$, again.
For both sets the subsidiary condition \eq{xev} yields $\lambda=\b{x}$ that implies
\bea
  \mu^2=\biggl(\frac{ N\pi}{2K} \biggr)^2
\eea
Therefore, the state centered at $x=\b{x}$ with minimal position uncertainty is the one 
with $N=1$,
\bea\label{minposun}
  \t{\t{\varphi}}_{\b{x}}(k_x)&=& \t{\t{\phi}}_{N=1} (k_x) =
    \sqrt{2a} e^{-ik_x \b{x}} \cos \biggl(\frac{k_x a}{2}\biggr)
\eea
with $a=\pi/K$ after normalization. What one has to check yet that this state is of finite energy. For a particle with the usual kinetic energy  $\frac{\h{p}_x^2}{2m}$ the integral
\bea
  \int_{-\infty}^\infty \frac{dp_x}{2\pi f(\alpha|p_x|)} p_x^2 \cos^2 \biggl(\frac{k_x(p_x) a}{2}\biggr)
\eea
should converge, which happens if the deformation function increases for $|p_x|\to \infty$
faster than $p_x^2$, a condition satisfied by the deformation functions $f=\exp (\alpha^2p_x^2)$, and $f= \exp (\alpha|p_x|)$ cited in 
Sect. \ref{intro}, except of the case with $f=1+\alpha^2p_x^2$, although the latter 
might be a good low-momentum approximation of some realistic deformation function. The function \eq{minposun} and its derivatives with respect to $k_x$ are bounded functions of $k_x$. Therefore, any potential energy which can be approximated with a sequence of polynomials will have finite expectation value in the state given by
Eq. \eq{minposun}. Therefore, we can consider the wavefunction in Eq. \eq{minposun}
for the particular class of the deformation functions
 as that of the physical state of a particle centered at $\b{x}$ with the minimal position uncertainty, indeed. It is an advantage of the wavevector representation as compared to the canonical momentum representation that the wavevector wavefunctions of the states maximally
 localized at various positions do not depend on the explicit form of the deformation 
function.

\section{Reconstruction of a continuous bandlimited potential from sampled values of Dirac-delta potential}\label{recconp}

First, we give 
 a unique definition of the sum $\sum_{n=-\infty}^\infty e^{iqna}$ which turns out to be useful for the determination of reconstructed bandlimited potentials. We settle the ordering of the terms in the sum via
\bea
  \sum_{n=-\infty}^\infty e^{iqna}                 
=\lim_{N\to \infty}\biggl(
\sum_{n=0}^N e^{-iqna} +\sum_{n=0}^N e^{inqa}-1 \biggr),\nn
\eea
i.e. in the manner that ensures the completeness of the eigenstates $\t{\t{\psi}}_{x_n}(k_x)=\sqrt{a} e^{-ik_x x_n}$ $(x_n=na+\theta,~~\theta\in \lbrack 0, a),~n\in \mathbb{Z})$ of an arbitrarily chosen 
self-adjoint extension $\h{x}_\theta$ of the coordinate operator $-i\partial_{k_x}$ in the 
 bandlimited Hilbert space $\c{H}$. 
 Then one finds
\bea\label{sumdef}
\lefteqn{
  \sum_{n=-\infty}^\infty e^{iqna+ iq\theta}                 }\nn
&=& \lim_{N\to \infty}e^{iq\theta}\biggl(
 \frac{ e^{-iq(N+1)a}-1}{e^{-iqa}-1}  + \frac{ e^{iq(N+1)a}-1}{e^{iqa}-1} -1\biggr)\nn
&=&
 \lim_{N\to \infty}e^{iq\theta}\frac{ \cos (Nqa) -\cos \lbrack (N+1)qa\rbrack }{1-\cos (qa)}
\nn
&=&
\pi qa e^{iq\theta} \lim_{N\to \infty}\biggl( \frac{ \cos (Nqa)}{\pi qa} +\frac{\sin (qa)}{1-\cos (qa)}
\frac{\sin (Nqa)}{\pi qa} \biggr)\nn
&=&
\pi qa \biggl( -1 +\frac{\sin (qa)}{1-\cos (qa)}\biggr) \delta (qa)\nn
&=&
\frac{2\pi}{a}\delta (q).
\eea
Let us emphasize that the definition of the sum does not depend on the choice of $\theta$, i.e. that of the self-adjoint extension of the coordinate operator.
Hence   the completeness relation of the coordinate eigenstates $|x_n^\theta\ra$ in 
the bandlimited Hilbert space $\c{H}$ takes the form
\bea
  \sum_{n=-\infty}^\infty \t{\t{\psi}^\theta}_{\!\!\!\!\!x_n}^* (k_x)
\t{\t{\psi}^\theta}_{x_n}(k_x')
&=&
a \biggl(  \sum_{n=-\infty}^\infty e^{i(k_x-k_x')(na+\theta) } \biggr) \nn
&=&
2\pi \delta (k_x-k_x') 
\eea
in the wavevector representation.

As to the next, we determine the bandlimited continuous potential $\b{V}(\b{x})$ reconstructed from discrete sampled values $\b{V}_n$ (with the notations of Sect.
\ref{smear}) of the Dirac-delta  like potential  $V(x)=V_0 a\delta(x)$.
Making use of the reconstruction formula in Eq.  \eq{potrec}, the sample \eq{didesa},
 and the rule to evaluate the sum like  in Eq. \eq{sumdef},
 one reconstructs the following continuous bandlimited potential,
\bea\label{didere}
\lefteqn{
  \b{V}(\b{x})     }\nn
&=& \frac{V_0 a^3}{2} \sum_{n=-\infty}^\infty 
\lbrack \Pi(x_n-\hf a)+\Pi(x_n+\hf a)\rbrack^2\Pi(\b{x}-x_n)\nn
&= &
 \frac{V_0 a^3}{2}   \sum_{n=-\infty}^\infty \int_{-K}^K \frac{dk_1}{2\pi}
\lbrack  e^{ik_1 (x_n-\hf a)}+ e^{ik_1(x_n+\hf a)}\rbrack\nn
&&\times
\int_{-K}^K \frac{dk_2}{2\pi} \lbrack  e^{ik_2 (x_n-\hf a)}+ e^{ik_2(x_n+\hf a)}\rbrack \nn
&&\times
\int_{-K}^K \frac{dq}{2\pi} e^{iq(\b{x}- x_n)}\nn
&=&
 \frac{V_0 a^3}{2}  \int_{-K}^K \frac{dk_1}{2\pi} \int_{-K}^K \frac{dk_2}{2\pi} 
 \int_{-K}^K \frac{dq}{2\pi}\lbrack  e^{-ik_1 \hf a}+ e^{ik_1\hf a}\rbrack\nn
&&\times\lbrack  e^{-ik_2 \hf a}+ e^{ik_2\hf a}\rbrack 
 e^{iq\b{x}}  \sum_{n=-\infty}^\infty e^{i(k_1+k_2-q)x_n}\nn
&=&
\frac{ V_0 a^2}{2(2\pi)^2} \int_{-K}^K dk_1 \int_{-K}^K dk_2
\lbrack  e^{-ik_1 \hf a}+ e^{ik_1\hf a}\rbrack\nn
&&\times
\lbrack  e^{-ik_2 \hf a}+ e^{ik_2\hf a}\rbrack
 \int_{-K}^K dq e^{iq\b{x}} \delta(k_1+k_2-q)\nn
&=&
 \frac{V_0a^2 }{2(2\pi)^2}\int_{-K}^K dk_1\int_{-K}^K dk_2\lbrack  e^{-ik_1 \hf a}+ e^{ik_1\hf a}\rbrack\nn
&&\times
\lbrack  e^{-ik_2 \hf a}+ e^{ik_2\hf a}\rbrack
e^{i(k_1+k_2)\b{x}} \nn
&&\times\biggl( \Theta(k_1+k_2+K)-\Theta(K-k_1-k_2)\biggr)\nn
&= &
\frac{ V_0a^2 }{2(2\pi)^2}\lbrack  I(\b{x})+J(\b{x})\rbrack
\eea
with
\bea
I(\b{x})
&=&
 \int_{-K}^0 dk_1 \int_{-K-k_1}^K dk_2
\lbrack  e^{-ik_1 \hf a}+ e^{ik_1\hf a}\rbrack\nn
&&\times
\lbrack  e^{-ik_2 \hf a}+ e^{ik_2\hf a}\rbrack
e^{i(k_1+k_2)\b{x}} ,\nn
J(\b{x}) &=&\int_0^K dk_1 \int_{-K}^{K-k_1} dk_2 \lbrack  e^{-ik_1 \hf a}+ e^{ik_1\hf a}\rbrack\nn
&&\times
\lbrack  e^{-ik_2 \hf a}+ e^{ik_2\hf a}\rbrack
e^{i(k_1+k_2)\b{x}} \nn
&=& I(-\b{x}).
\eea 
Then a somewhat lengthy but straightforward calculation yields
\bea\label{didere2}
\b{V}(\b{x})&=& V_0\frac{ a^2}{(2\pi)^2\lbrack\b{x}^2-(a/2)^2\rbrack}
\biggl( \frac{ a^2}{ \b{x}^2-(a/2)^2}\nn
&& -\frac{ 2a\b{x} \sin (K\b{x}) }{ \b{x}^2-(a/2)^2}
 + \frac{4\b{x}}{a} \sin K\b{x}  -\pi \cos K\b{x} \biggr).\nn
\eea
This function turns out to be a unique even function of $\b{x}$ in all of the various $\theta$
 sectors. It takes the typical values $\b{V}(0)=V_0\frac{4+\pi}{\pi^2}\approx 0.72 V_0$, $\b{V}(\pm \hf a)=V_0\frac{ 1+ 5(\pi/4)^2}{\pi^2}\approx 0.41 V_0$ and falls off rapidly outside the interval $\b{x}\in \lbrack -\hf a,\hf a\rbrack$ in an oscillatory manner.

\end{document}